\documentclass[12pt]{article}
\usepackage{xtex}
\usepackage{graphicx}
\newcommand{\beq}{\begin{equation}}
\newcommand{\eeq}{\end{equation}}
\newcommand{\beqa}{\begin{eqnarray}}
\newcommand{\eeqa}{\end{eqnarray}}
\newcommand{\Gg}{\ensuremath{\bar{G}}}
\newcommand{\half}{\ensuremath{\frac{1}{2}}}
\newcommand{\ttx}{\ensuremath{\bar{T}^{tx}}}
\newcommand{\txx}{\ensuremath{\bar{T}^{xx}}}
\newcommand{\eq}[1]{Eq.~(\ref{#1})}
\newcommand{\eqs}[2]{Eqs.~(\ref{#1}) and (\ref{#2})}
\def\be{\begin{equation}}
\def\ee{\end{equation}}
\def\ba{\begin{eqnarray*}}
\def\ea{\end{eqnarray*}}

\def\imq{\mbox{Im}~q}
\def\imp{\mbox{Im}~p}

\begin{document}

\begin{titlepage}

\setcounter{page}{1} \baselineskip=15.5pt \thispagestyle{empty}

%\begin{flushright}
%Local preprint number here-\\ % hep-th/yymmnnn\\ % no longer needed?
%\end{flushright}
%\vfil

\begin{center}
{\LARGE Shock waves in strongly coupled plasmas}
\end{center}
\vspace{\bigskipamount}

\begin{center}
{\large Sergei Khlebnikov, %\footnote{skhleb@physics.purdue.edu}
 Martin Kruczenski %\footnote{markru@purdue.edu} 
 and Georgios Michalogiorgakis%\footnote{gmichalo@purdue.edu}
}
\end{center}

\begin{center}
\textit{Department of Physics, Purdue University \\ 
525 Northwestern Avenue, West Lafayette, IN 47907} 
\end{center} %\vfil
\vskip .01 in
\begin{center}
{\tt skhleb@physics.purdue.edu \quad markru@purdue.edu \quad gmichalo@purdue.edu } 
\end{center}
\vfil
%\vskip 3 in 

%\begin{center}
 \noindent 
Shock waves are supersonic disturbances propagating in a fluid and giving rise to dissipation and drag. Weak shocks, i.e., those of small amplitude, 
can be well described within the hydrodynamic approximation. On the other hand, strong shocks are discontinuous within hydrodynamics
and therefore probe the microscopics of the theory. In this paper we consider the case of the strongly coupled $\mathcal{N}=4$ plasma
whose microscopic description, applicable for scales smaller than the inverse temperature, is given in terms of gravity in an 
asymptotically $AdS_5$ space. In the gravity approximation, weak and strong shocks should be described by smooth metrics
with no discontinuities. For weak shocks we find the dual metric in a derivative expansion and for strong shocks we use linearized
gravity to find the exponential tail that determines the width of the shock. In particular we find that, when the velocity of the fluid relative to the shock
approaches the speed of light $v\rightarrow 1$ the penetration depth $\ell$ scales as $\ell\sim (1-v^2)^{1/4}$. We compare the results with
second order hydrodynamics and the Israel-Stewart approximation. Although they all agree in the hydrodynamic regime of weak shocks, we show
that there is not even qualitative agreement for strong shocks. 
For the gravity side, the existence of shock waves implies that there are disturbances of constant shape propagating on the horizon of the dual black holes. 

%\end{center}
 
\vfil

\end{titlepage}
% ----/\----\/----/\----\/----/\----\/----/\----\/----/\----\/----/\----\/----
% ----/\----\/----/\----\/----/\----\/----/\----\/----/\----\/----/\----\/----

%\begin{document}

\tableofcontents

\clearpage

\section{Introduction and summary}\label{INTRO}

 The AdS/CFT correspondence \cite{Maldacena:1997re,Gubser:1998bc,Witten:1998qj,Aharony:1999ti} realizes in practice the idea of describing gauge theories 
in terms of strings \cite{'tHooft:1973jz} and provides a new tool to study gauge theories at strong coupling.
Recently, motivated by experimental and theoretical work on heavy ion collisions at RHIC, there was a wave of interest in using AdS/CFT to study the 
dynamics of strongly coupled plasmas, for a review see \cite{Son:2007vk,Shuryak:2008eq,Schafer:2009dj,Gubser:2009md,Gubser:2009sn} and references therein. The 
main initial focus was on the low viscosity of the fluid and the drag force experienced by a quark moving through the plasma. More generically, 
in conformal plasmas, such as the one studied in AdS/CFT, the hydrodynamic description of the theory is valid up to length scales of order of the inverse 
temperature. Below that scale standard hydrodynamics breaks down, and we need to resort to the dual gravity description, which does not break down until 
the (much smaller) bulk string scale is reached. Alternative descriptions that are not based
directly on microscopics, such as the Israel-Stewart theory, can be tested by comparison with
the microscopic theory provided by the dual gravity.

In the present paper, we focus on a particular phenomenon in fluid dynamics, 
namely, shock waves. These are supersonic disturbances that propagate in the fluid and are typically produced by an object moving supersonically in the fluid or by 
a localized release of energy such as in explosions or collisions. In ideal fluids they are described by a surface of discontinuity where the normal 
velocity and the pressure have a jump. In the frame where the shock is at rest, the fluid goes from supersonic to subsonic, with the kinetic energy of 
the fluid converted into pressure and heat. This process is irreversible and generates entropy. In fact, it is the only dissipative
mechanism in the case of strictly ideal fluids. In practice, of course, the fluid is never ideal so at small scales the viscosity also causes 
dissipation. If the shock wave is produced by an object moving in the fluid it also generates drag. Away from the shock the hydrodynamic approximation 
is good and allows to compute the jumps in velocity and pressure by using the conservation of energy (and other charges) across the shock. However 
in the region of discontinuity the gradients are in general large and a microscopic theory is necessary to have an appropriate description. In fact this breakdown 
of hydrodynamics is the reason why dissipation occurs in shock waves even for ideal fluids. 
For that reason, shock waves are not only phenomenologically 
interesting but, on the theoretical side, can be considered as a useful probe of the system. In practice, they are usually studied in non-relativistic systems and in fluids such
as air and water. The relativistic case is of interest in astrophysics and in particular in heavy ion physics, which is the system closest to the one 
studied in this paper. 

 In the context of heavy ion physics there are different situations where shock waves can appear. One is during the period of the creation of the Quark 
Gluon Plasma. A simple model of particle production which gives reasonable results is Landau's
hydrodynamic model \cite{Landau:1953gs}. This model assumes that the fluid 
thermalizes instantaneously and then evolves according to ideal hydrodynamics. All the entropy is created at the initial stage as a result
of the thermalization process. In a purely ideal hydrodynamical model the entropy can be generated only from shock waves which would therefore play an
important role. However, initially the quark gluon plasma is far from equilibrium and therefore a hydrodynamical approximation might not be valid. 

  Within AdS/CFT entropy production in a heavy ion collision has been studied in \cite{Gubser:2008pc} where two shock waves moving at the speed of light 
collide creating trapped surfaces.  Also, \cite{Avsar:2009xf} study the problem of reconstructing the bulk metric from the boundary.  Several other aspects 
are examined in \cite{Beuf:2009mk,Albacete:2008vs}.  Related work on gravitational shock waves appears in \cite{Aichelburg:1970dh,Dray:1984ha,Hotta:1992qy,Sfetsos:1994xa,Podolsky:1997ni,Horowitz:1999gf,Emparan:2001ce,Arcioni:2001my,Horowitz:2009fw,Horowitz:2009pw}.    

Another instance where shock waves might appear in heavy ion collisions is the process where a hard parton moves through the 
plasma.  The hard parton can be  a heavy quark, a meson or a gluon.  
Calculations in AdS/CFT \cite{Friess:2006fk,Gubser:2007xz,Gubser:2007ga,Chesler:2007an,Chesler:2007sv} show that there is a Mach cone created.  
That is, the energy density of the wave is concentrated in a angle close to the Mach angle.  This is highly suggestive of the formation of a shock wave that 
cannot be described by the linearized approximations used in previous calculations.  The real space profile of the stress energy tensor of a 
moving quark, calculated in \cite{Gubser:2007xz} also supports the idea.  Independently from AdS/CFT the possibility of shock waves in heavy ion collisions has been explored both theoretically and experimentally in \cite{Scheid:1974zz,Baumgardt:1975qv,Gutbrod:1989wd,Gutbrod:1989gh,Adams:2003kv,Adare:2008qa,Wang:2004kfa,Adams:2005ph,Adler:2005ee,Ulery:2005cc,Ajitanand:2006is,Adare:2008cqb,Stoecker:2004qu,Ruppert:2005uz,Koch:2005sx,CasalderreySolana:2004qm,Bouras:2009nn,Bouras:2010nt}.     

\section{Shock waves in hydrodynamics}\label{SHHYDRO}

\begin{figure}
 \begin{center}
\includegraphics[height=6cm,width=9cm]{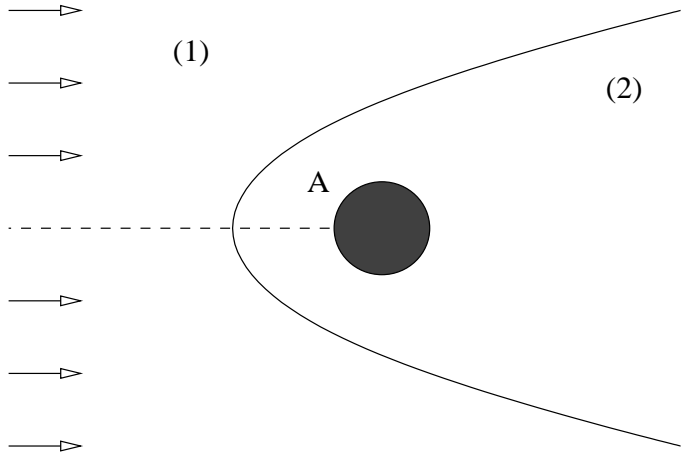}
 \end{center}
 \caption[]{A supersonic flow reaches a fixed object creating a stationary shock wave. In heavy ion physics this scenario might be realized by a heavy 
parton moving through the plasma as suggested by the existence of a Mach cone.}
\label{ShockSketch}
 \end{figure}

 Hydrodynamics can be used to study the region away form the shock where the gradients are small. For weak shocks, i.e., those of small amplitude, a 
hydrodynamic approximation that includes viscosity resolves the discontinuity and provides a smooth description of the shock wave. This can be 
corrected at higher orders in gradients if so desired. In this section we start by considering the ideal fluid case for arbitrary shocks and then the higher 
order hydrodynamic approximation for weak shocks. We also analyze the Israel-Stewart theory to compare later with the microscopic results. 

\subsection{Ideal Hydrodynamics}\label{IDEALHYDRO}

 Ideal hydrodynamics is valid far from the shock and determines the main properties of shock waves. These are well known \cite{LL} and we describe them 
in this subsection as applied to our particular fluid.  In relativistic ideal hydrodynamics the energy momentum tensor is given by
\eqn{IDEALHYD}{
T^{\m\n}=p \h^{\m\n} +\left(p+\r \right) u^{\m} u^{\n} \;,\quad 
} 
where $p$ is the pressure, $\rho$ the energy density and $u^{\mu}$ the four velocity of the fluid. 
The equations of motion are given by conservation of $T^{\mu\nu}$,
\beq
\p _{\m} T^{\m\n}=0\; ,
\eeq
and the equation of state, which for the strongly coupled $\mathcal{N}=4$ plasma is dictated by conformality and reads\footnote{We follow the conventions of \cite{Bhattacharyya:2008jc} with regard to the normalization of the stress energy tensor.  That is our stress tensor is related to the conventionally defined one by $T^{\m\n}_{our}=\frac{8 \pi^2}{N^2}T^{\m\n}$.  In gravity this is reflected in $T^{\m\n}_{our}= 16\pi G_{N}^{[5]} T^{\m\n}$ where $G_{N}^{[5]}$ is the five dimensional Newton's constant.}
\eqn{EOSN}{
\r=3p =3 (\pi T)^4\;.
} 
For a shock wave with a planar front, located for convenience  at $x=0$, both $T^{tx}$ and $T^{xx}$ are constant throughout the fluid:
\eqn{Ttxideal}{
T^{tx}=4p u^{t}u^{x}=4p\frac{v}{1-v^2}\;,  
}
\eqn{Txxideal}{
T^{xx}=p\left(1+4u_{x}^{2}\right)=p\frac{1+3v^2}{1-v^2}\;.
}
 The possibility of a shock wave arises because there are different values of pressure and velocity that give the same value of $T^{tx}$ and $T^{xx}$.  
For example, the ratio $T^{tx}/T^{xx}$ depends only on the velocity and, as is seen from the plot in fig.\eno{FigIdealShock} different velocities can 
give the same ratio $T^{tx}/T^{xx}$. More precisely, if we take that for $x \rightarrow -\infty$, $v=v_{1}$ and $T=T_{1}$ and 
for $x\rightarrow \infty$, $v=v_2$ and $T=T_2$  we can ensure that $T^{tx}$ and $T^{xx}$ are constant by imposing 
\eqn{GTOV2}{
v_{2}= \frac{1}{3v_{1}} \;,\quad p_{2}= p_{1} \frac{9v_{1}^2-1}{3(1-v_{1}^2)}\;,\quad T_{2} =T_{1} \left(\frac{9v_{1}^2-1}{3(1-v_{1}^2)}\right)^{1/4}\;.
}     
 A particular case is $v_1=v_2=v_s$ where $v_s=\frac{1}{\sqrt{3}}$ is the speed of sound. In that case the jumps in $v$ and $T$ vanish. In general we have
\beq
 1>v_1>\frac{1}{\sqrt{3}}>v_2>\frac{1}{3}
\eeq
as can be seen from the relation $v_1 v_2=\frac{1}{3}$. 
In particular, if $v_2 = 1/3$ on one side of the shock, the fluid on the other side moves at
the speed of light. In principle, the conservation laws allow one to switch the values of the
velocity between the front and back of the shock, i.e., take $v_1<v_s$ and $v_2>v_s$, but such solution would convert thermal energy into kinetic energy violating the second law of thermodynamics. 
In figure \eno{FigPressIdeal} we show the ratio of the pressures on the two sides.  The supersonic side of the shock has a lower pressure which goes to zero 
when $v_{1}$ approaches one.      
\begin{figure}
 \begin{center}
\includegraphics[height=4cm,width=6cm]{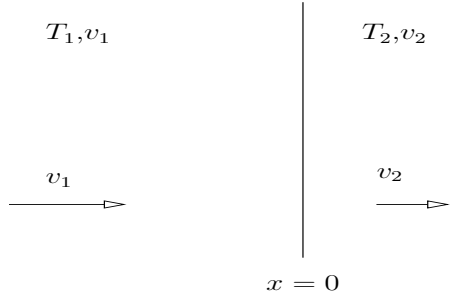}
 \end{center}
 \caption[]{Sketch of a shock wave in the rest frame of the interface.  For ideal hydrodynamics there is a discontinuity at $x=0$.  Including the higher order terms in the expression of the stress energy tensor resolves the discontinuity.  The conventions of the paper are that the fluid moves to the right. The left hand side is supersonic and the right hand side is subsonic since the opposite configuration violates the second law of thermodynamics.}
\label{ZeroShockFig}
 \end{figure}
For an object moving through the fluid this has the effect of changing the pressure that the object experiences. Indeed, defining the projector perpendicular to $u^\mu$ as 
$P^{\mu\nu}=\eta^{\mu\nu}+u^\mu u^\nu$ we obtain, from eq.(\ref{IDEALHYD}):
\beq
 P_{\mu\nu} \partial_\alpha T^{\alpha \nu} = 0, \ \ \ \Rightarrow \ \ \ \ \partial_\mu T + u^{\nu} \partial_\nu \left(T u_{\mu}\right)=0
\eeq
For a stationary solution, taking the $\mu=0$ component we obtain:
\beq
 u^{i} \partial_{i} \left(\gamma T\right) =0
\eeq
after replacing $u^0=\gamma$, where $\gamma$ is the Lorentz factor. We see that $\gamma T$ is conserved along a streamline. In particular for a situation as in fig.\ref{ShockSketch}
we can follow a streamline from infinity to the point $A$ and obtain, with and without a shock wave, the pressure:
\beq
p_A^{\mbox{(shock)}} = \gamma_{2}^4\, p_{2}=\frac{27v_{\infty}^4}{(9v_\infty^2-1)(1-v_\infty^2)} p_{\infty}, \ \ \ \ \  p_A^{\mbox{(no shock)}} = \gamma_{\infty}^4\, p_{\infty}= \frac{ p_{\infty}}{(1-v_{\infty}^2)^2}\;,
\eeq
where we have used eq.(\ref{EOSN}) and the matching conditions (\ref{GTOV2}) for the pressure and velocity across the shock wave.

Another interesting aspect of shock waves is the entropy production that is associated with them.  In ideal hydrodynamics the entropy current is given by
\eqn{SCURR}{
s_{\m}= 4 \pi^4 T^3 u_{\m}\;,
}
where we have chosen the normalization to give the entropy density of a fluid at rest.  The difference between $s^0$ on the two sides of the shock is given by
\eqn{DSIDEAL}{
\D s^{0} &= s^{0}_{subsonic}-s^{0}_{supersonic}= 4\pi^4 \left(T_{2}^{3} \sqrt{1+u_{2}^{2}}-T_{1}^{3}\sqrt{1+u_{1}^{2}}\right) \; , \cr 
\D s^{0} &= \frac{4 \pi^4 T_{1}^3}{\sqrt{1-v_{1}^2}} \left(3^{1/4}v_{1}\left(\frac{9v_{1}^2-1}{1-v_{1}^2}\right)^{1/4} -1\right) \;.
}
The entropy production comes from the difference of the two fluxes 
\eqn{FNTFLUX}{
\D s^{x} =  s^{x}_{subsonic}-s^{x}_{supersonic}=\frac{4\pi^2 T_{1}^{3}}{\sqrt{1-v_{1}^2}} \left(3^{-3/4}\left(\frac{9v_{1}^2-1}{1-v_{1}^2}\right)^{1/4} -v_1\right)\;.
}
Kinetic energy from the supersonic side is transformed into thermal energy in the subsonic side and this creates entropy. Notice also that the ideal hydro 
is correct far from the shock, so this calculation gives the correct entropy production even if the fluid is not ideal. 

\begin{figure}
 \begin{center}
\includegraphics[height=6cm,width=9cm]{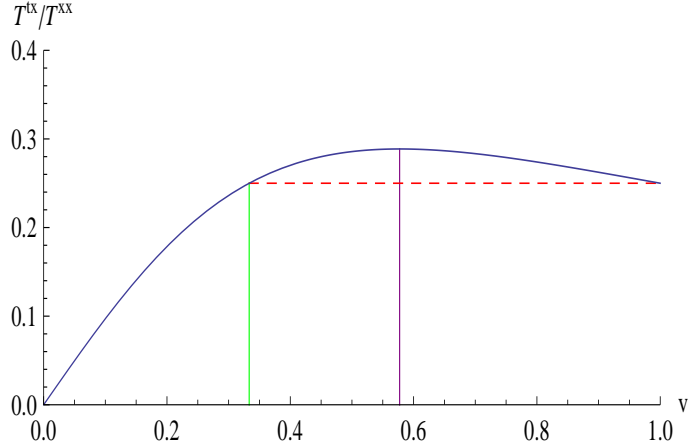}
 \end{center}
 \caption[]{The ratio $T^{tx}/T^{xx}$ for different velocities.  For  $v_{1}>\frac{1}{3}$  the equation $T^{tx}/T^{xx}=const.$ has two solutions and hence allows for a jump in velocity.
 When $v_{1}=\frac{1}{\sqrt{3}}$  there is no jump.}
\label{FigIdealShock}
 \end{figure}

\begin{figure}
 \begin{center}
\includegraphics[height=6cm,width=9cm]{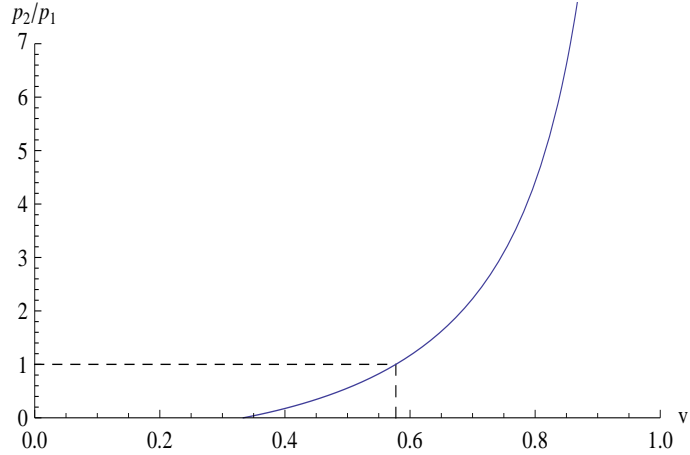}
 \end{center}
 \caption[]{The ratio of pressure for the two sides of a shock wave as a function of the incident velocity $v_{1}$.  When $v_{1}<1/3$ there is no shock. For $v_{1}<\frac{1}{\sqrt{3}}$,$v_{2}>\frac{1}{\sqrt{3}}$ the pressure and the temperature on the supersonic side $(2)$ are smaller than those on the subsonic side. Kinetic energy from the supersonic side is transformed to thermal energy on the subsonic side thereby increasing the entropy.  The entropy production in this process is discussed further in section \eno{SECORDERHYD}.  }
\label{FigPressIdeal}
 \end{figure}

\subsection{Viscous Hydrodynamics}\label{VISCHYDRO}

The stress-energy tensor for relativistic hydrodynamics can be organized in a series expansion in powers of $\frac{1}{LT}$ where $L$ is a typical length scale over which the four-velocity changes and $T$ is the temperature.  In such an expansion the first order term is the viscous term.  Since the plasma we are interested in is conformal, its bulk viscosity is zero. The shear viscosity is given by $\frac{\h}{s} =\frac{1}{4\pi}$ \cite{Policastro:2001yc}.  The stress energy tensor to the first order is given by
\eqn{STRESSF}{
T^{\m\n}= (\pi T)^{4}\left( \h^{\m\n} +4u^{\m}u^{\n}\right) -2(\pi T)^{3} \s ^{\m\n}\;,
} 
where
\eqn{SIGMAPIE}{
P^{\m\n}= \h^{\m\n} +u^{\m}u^{\n} \;, \quad \s^{\m\n}=P^{\m\a}P^{\n\b}\p_{(\a}u_{\b)}-\frac{1}{3}P^{\m\n}\p_{\a}u^{\a} \;.
}
Notice that
\eqn{DEFTU}{
T^{\m\n}u_{\n} = -3(\pi T)^4 u^{\m}\;.
} 
 which can be taken as the definition of $u^{\mu}$ and $T$, namely $u^{\mu}$ is a time like eigenvector of $T^{\m\n}$ whose eigenvalue is $-3(\pi T)^4$. 
This definition can be used at any order in the hydrodynamic expansion and is sometimes called the Landau frame.  

 Now we would like to see how a weak shock wave is resolved if the effects of viscosity are included. We consider a flow where the four-velocity and temperature are functions only of $x$:
\eqn{FLOWX}{
T=T(x)\;,\quad u_{\m} = (-\sqrt{1+u(x)^2},u(x),0,0)\;.
}   
As before, conservation of the energy-momentum tensor implies that the components $T^{tx}$ and $T^{xx}$ are constant throughout the fluid. They are now given by
\eqn{GOTTVISC0X}{
T^{tx}=4(\pi T)^4 u(x)\sqrt{1+u(x)^2}-\frac{4}{3} (\pi T)^3 u(x)\sqrt{1+u(x)^2} u'(x) =\ttx \;,
}
\eqn{GOTTVISCXX}{
T^{xx}=(\pi T )^4 (1+4u(x)^2) -\frac{4}{3}(\pi T)^3 (1+u(x)^2)u'(x) =\txx \;.
}
The asymptotic behavior of $T,u(x)$ determines the constants $\ttx,\txx$.  Conversely, in terms of $\ttx,\txx$ and $u(x)$ the temperature is given by 
\eqn{GOTT}{
\pi T = \frac{1}{3^{1/4}} \left(\ttx\frac{\sqrt{1+u(x)^2}}{u(x)}-\txx \right)\;,
}
which is in fact valid to all orders in the hydrodynamic expansion since it follows from the definition (\ref{DEFTU}).  

Let us suppose that asymptotically on the supersonic side the temperature and velocity approach constants: 
\eqn{BOUNDARYC}{
T_{1} = \lim_{x\rightarrow- \infty} T(x)=T_{(0)}\left(1-\frac{\sqrt{2}}{3}u_{\infty}\right) \;,\quad u_{1}= \frac{1}{\sqrt{2}} + u_{\infty} = \lim_{x\rightarrow -\infty}u(x)\;.
} 
Four-velocity $\frac{1}{\sqrt{2}}$ corresponds to
the speed of sound in a conformal plasma. 
The remaining equation $T^{tx}=\ttx$ gives a 
differential equation for $u(x)$.  Since first-order hydrodynamics is valid only for weak shocks we expand 
in a power series in $u_{\infty}$. It then becomes clear that it is useful to define a new variable
\eqn{XIDEF}{
\xi =\frac{4\pi T_{(0)} u_{\infty}}{3} x\;,
}
in terms of which we have
\eqn{USERIES}{
u(\xi)=\frac{1}{\sqrt{2}}+ u_{\infty} \d u_{(1)}(\xi)+ u_{\infty}^{2}\d u_{(2)}(\xi) + \ldots \;.
}
The equations of motion imply that $\d u_{(1)}(\xi)$ satisfies the equation (primes denote derivation with respect to $\xi$)
\eqn{U1EQ}{
\d u_{(1)}'=\d u_{(1)}^2-1\;,
}
with solution 
\eqn{GOTU1}{
\d u_{(1)} =\tanh(-\xi)\;.
}
This has the same form as the solution for a weak shock in nonrelativistic hydrodynamics \cite{LL}, and indeed
\eq{U1EQ} coincides with the first integral of the Burgers equation, familiar in that context.

As we already noted, in ideal hydrodynamics one can freely exchange the two sides of the shock, so that in 
the rest frame of the shock the fluid's velocity may change either from subsonic to supersonic or from 
supersonic to subsonic.
However, the existence of friction in the first order hydrodynamics breaks this symmetry and only the latter solution is allowed.

The approach to the asymptotic values of $T$ and $u$ is described by 
$T(x)\sim T_{asympt.}+e^{iqx}\, \d T $, $u(x)\sim u_{asympt.} +e^{iqx}\, \d u$, where $T_{asympt.}$ is either $T_1$ or 
$T_2$, depending on which of the asymptotic regions we are looking at. Expanding the equations for conservation of 
energy and momentum to the first order in $\d T, \d u$ provides us with a system of two equations with two unknowns, $\d T, \d u$.  Demanding that there is a non-zero solution determines $q$ to be\footnote{Notice that this gives real exponential that decay away from the shock.}
\eqn{GOTQF}{
\frac{iq}{\pi T} =\frac{\sqrt{1-v^2}}{v}(3v^2-1)\, 
} 
where we use $u_{asympt.}=\frac{v}{\sqrt{1-v^2}}$. For weak shocks, we expand \eno{GOTQF} around the speed of sound $v\sim \frac{1}{\sqrt{3}}+ \d v$ to obtain
\eqn{GOTQFEXP}{
\frac{iq}{\pi T} = 2\sqrt{6} \d v\; ,
}
which agrees with the explicit solution \eno{GOTU1}. For strong shocks, for which $|q|$ is not small in comparison
with $T$, there is no reason to expect \eno{GOTQF} to be a good approximation. We compare it with
other approximations in subsequent sections.

\subsection{Second order hydrodynamics and Israel-Stewart theory}\label{SECORDERHYD}

Let us now consider how shock waves are resolved in second order hydrodynamics and in Israel-Stewart theory.  For the $\mathcal{N}=4$ plasma, the stress energy tensor has been computed to second order in \cite{Baier:2007ix,Bhattacharyya:2008jc} (see also \cite{Natsuume:2007ty}), and is given by
\eqn{STRESSSEC}{
T^{\m\n}=& (\pi T)^4 \left(\h^{\m\n}+4u^{\m}u^{\n}  \right)-2(\pi T)^3 \s^{\m\n} \cr
                  & +(\pi T)^2 \left((\ln 2) T_{2a}^{\m\n} +2 T_{2b}^{\m\n} +(2-\ln 2)\left(\frac{1}{3}T_{2c}^{\m\n} +T_{2d}^{\m\n}+T_{2e}^{\m\n}\right)  \right)\;,
}
where 
\eqn{SIGDEF}{
\s^{\m\n} = P^{\m\a}P^{\n\b}\p_{(\a} u_{\b)}-\frac{1}{3} P^{\m\n} \p_{\a} u^{\a}\;, \quad \mathcal{D} = u^{\a}\p_{\a}
}

\eqn{T2ADEF}{
T_{2a}^{\m\n} =\e ^{\a\b\g (\m}\s^{\n)}_{\g}u_{\a}l_{\b}\;, T_{2c}^{\m\n} = \p_{\a} u^{\a} \s ^{\m\n}\;,
}
\eqn{T2BDEF}{
T_{2b}^{\m\n}=\s^{\m\a}\s^{\n}_{\a} -\frac{1}{3}P^{\m\n}\s^{\a\b}\s_{\a\b}\;,
}
\eqn{T2DDEF}{
T_{2d}^{\m\n}=\mathcal{D}^{\m}\mathcal{D}u^{\n} -\frac{1}{3}P^{\m\n} \mathcal{D} u^{\a} \mathcal{D} u_{\a}\;, 
}
\eqn{T2EDEF}{
T_{2e}^{\m\n}= P^{\m\a}P^{\n\b} \mathcal{D} \left( \p_{(\a}u_{\b)}\right)-\frac{1}{3}P^{\m\n} P^{\a\b} \mathcal{D} \left(\p_{\a}u_{\b} \right)\;,
}
\eqn{LDEF}{
l_{\m}=\e_{\a\b\g\m}u^{\a}\p^{\b}u^{\g}\;,
}
 We follow the conventions of \cite{Bhattacharyya:2008jc} where $\e_{0123}=1$ and the brackets denote symmetrization. The $T^{0x}$ and $T^{xx}$ components of the stress tensor are again constant but now given by
\eqn{GOTT0X2}{
T^{0x}= & u(x) \sqrt{1+u(x)^2}\Bigg(4(\pi T(x))^4 -\frac{4}{3}(\pi T(x))^3+\cr     
               &(\pi T(x))^2 \left(\frac{2}{9}(4-\ln 2)u'(x)^2+\frac{2}{3}u(x)u''(x) \right) \Bigg) \;,
} 
\eqn{GOTTXX2}{
T^{xx}=& (\pi T(x))^4\left(1+4u(x)^2\right) -\frac{4}{3}(1+u(x)^2)(\pi T(x))^3 +\cr
              &(\pi T(x))^2 (1+u(x)^2)\left(\frac{2}{9}(4-\ln 2)u'(x)^2+\frac{2}{3}u(x)u''(x) \right)\;.
}
 First, we carry out linear analysis near the asymptotics at $x\to \pm \infty$, where we expect 
\eqn{SECANSATZ}{
T(x)= T_{asympt.} +\d T e^{iqx}\;,\quad u(x)= u_{asympt.} +\d u e^{iqx}\;.
}  
Keeping only linear terms in $\d T,\d u$ we solve 
\eqn{EOMINT}{
T^{tx}=\ttx\;,\quad T^{xx}= \txx
}
where $\ttx,\txx$ are given by the asymptotic values 
\eqn{GOTCS}{
\ttx= 4(\pi T_{asympt.})^4 u_{asympt.}\sqrt{1+u_{asympt.}^2} \;,\quad \txx=(\pi T_{asympt.})^2 \left(1+4 u_{asympt.}^2 \right)\;.
 }
The 2 by 2 linear system for $\d u, \d T $ has a nonzero solution only if its determinant is zero.  This condition determines $q$ to be 
\eqn{GTOQSEC}{
\frac{iq}{\pi T_{asympt.}}= \frac{\sqrt{1-v^2} \left(1-\sqrt{1-2 \left(3 v^2-1\right) (2-\log (2))}\right)}{v (2-\log
   (2))}\;,
}
where $v= \frac{u_{asympt.}}{\sqrt{1+u_{asympt.}^2}}$. This is intended as an improvement on
the first-order formula \eno{GOTQF}. Note that the argument of the square root becomes negative for velocities greater than 
\eqn{V2FAIL}{
v > \sqrt{\frac{5-2 \log (2)}{12-6 \log (2)}} \approx0.678871\;.
} 
This indicates that, not unexpectedly, computations in second order hydrodynamics should not be trusted beyond the weak shock regime, i.e., beyond velocities close to the speed of sound, $|v-v_{s}| \ll v_s$. Note that the velocity \eno{V2FAIL} is different from the velocity of discontinuity propagation in second order hydrodynamics and Israel-Stewart theory 
\eqn{VDISC}{
v_{disc}= \sqrt{\frac{1}{2(2-\ln 2)}}\approx 0.618546\;.
}

Next, we solve for the shock solution for speeds that are close to the speed of sound.  To this end we have to use the solution of first order hydrodynamics \eno{GOTU1} and expand to second order in $u_{\infty}$, the difference 
between the actual asymptotic speed and the speed of sound. Using \eno{USERIES} we now solve 
\eqn{SECEOMS}{
T^{tx}=\ttx\;,\quad T^{xx}=\txx\;.
}   
Again the temperature field is given by \eno{GOTT} and the second term in the expansion of the velocity must satisfy 
\eqn{DU2}{
\d u_{2}'= \frac{1}{3} \Bigg(\d u_{1}  \Big(&\sqrt{2} \d u_{1}
   (1-(\log 4-2) \d u_{1})+\cr &+ 6 \d u_{2} +\sqrt{2} (\log 4-7)\Big)+4 \sqrt{2}\Bigg)
}
where, as before, the derivatives are with respect to $\xi= \frac{4\pi}{3} u_{\infty} T_{1} x$.  The solution is given by
\eqn{GOTU2}{
\d u_{2} (\xi)= \frac{1}{6}\left[4\sqrt{2}(1-\ln 2) \frac{\ln\cosh\xi}{\cosh^2\xi}
             +5\sqrt{2}\left(\tanh^2\xi+\tanh\xi+\frac{\xi}{\cosh^2\xi}\right)\right] \;.
}
This solution agrees with the one derived in gravity in section \eno{SHOCKGR}, using the prescription of \cite{Bhattacharyya:2008jc}. It also agrees with the linear analysis carried out above. Indeed, we can determine $q$ 
through
\eqn{GOTQ2}{
i q_{\pm}=\lim_{x\rightarrow \pm \infty} \frac{d^2 u/dx^2}{du/dx}
}
and compare them to the values following from \eno{GTOQSEC}.
Since this derivation is identical to the one we use in section \eno{FLUIDGR} we omit it here.

Finally, we examine the entropy production for this solution.  The entropy current has been derived in \cite{Loganayagam:2008is,Bhattacharyya:2008xc,Baier:2007ix}.  For second order hydrodynamics the current and the entropy production are given by 
\eqn{GOTSOENTRC}{
s^{\m} =4\pi \h u^{\m} - \frac{\t_{\pi} \h}{4T}\s^{\k\n}\s_{\k\n}u^{\m} \;,
}
\eqn{GOTSOENTRP}{
 \mathcal{\p}_{\m} s^{\m} = \frac{\h}{2T} \s^{\m\n}\s_{\m\n} \;.
}  
It is easy to check that the solution \eno{GOTU2} satisfies \eno{GOTSOENTRC}-\eno{GOTSOENTRP}.  Interestingly, as seen in figure \eno{EntrProd}, the entropy production is larger on the supersonic side of the wave. 
\begin{figure}
 \begin{center}
\includegraphics[height=6cm]{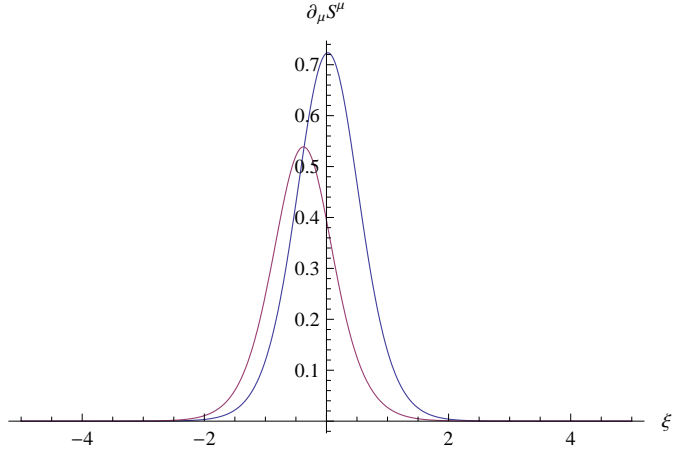}
 \end{center}
 \caption[]{Entropy production as determined by the divergence of the first order entropy current $s^{\m} =4\pi \h u^{\m}$ is plotted 
versus $\xi= \frac{4u_{\infty} x \pi T }{3}$.  For the first order solution \eno{GOTU1} (higher curve) the production is symmetric with respect to the front.  For the second order solution \eno{GOTU2} (lower curve) the entropy production is slightly displaced towards the supersonic side of the jet. The normalization of the vertical axis is such that the difference $\D s _{x}$ for the two sides for the first order solution is 1. Notice that the two curves do not have to integrate to the same number, since the asymptotic velocity and temperature differ for the first 
and second-order solutions, cf. Eqs.~\eno{GOTU1} and \eno{GOTU2}.}
\label{EntrProd}
 \end{figure}     

Similarly we can examine the asymptotic tail of shock waves in the Israel-Stewart theory \cite{Israel:1976tn,Israel:1979wp}. This is a theory originally proposed to cure the instantaneous propagation of discontinuities in first order relativistic hydrodynamics.  A new tensor $\pi_{\m\n}$ is introduced that parametrizes the departure from the ideal fluid:
\eqn{PIDEF}{
T^{\m\n} = p\left(\h^{\m\n}+4u^{\m}u^{\n}\right) +\pi^{\m\n}\;.
}  
The tensor $\pi^{\m\n}$ is connected to the velocity and temperature fields by 
\eqn{PICONST}{
\pi^{\m\n} +\t_{\pi}u^{\l}\mathcal{D}_{\l}\pi^{\m\n} =-2\h \s^{\m\n} +\t_{\oo}\left(\oo^{\m}_{\l}\pi^{\l\n}+\oo^{\n}_{\l}\pi^{\m\l}\right)\;,
}
where $\mathcal{D}_{\l}$ is the so called conformal derivative\footnote{For a definition of $\mathcal{D}_{\l},\oo^{\m\n}$ 
and a comparison between the Israel-Stewart theory and second order hydrodynamics for conformal plasmas one can consult \cite{Loganayagam:2008is}} and $\tau_\pi$ 
(or alternatively $\tau_{\oo}$) is a parameter the value of which has to be determined from microscopics. One often uses the rescaled, dimensionless parameters
$\bar{\h}$ and $\bar{\t}_\pi$ defined by
\eqn{RESCPARA}{
\h = \bar{\h} (\pi T)^3\;,\quad \t_{\pi} =\frac{\bar{\t}_{\pi}}{\pi T}\; .
}
For the $\mathcal{N}=4$ superconformal plasma 
\eqn{SCFTIS}{
\bar{\h}=1, \quad \bar{\t}_{\pi}= \frac{2-\ln 2}{2} \;.
}
Alternatively, one may initially leave these parameters undetermined and then choose them to fit specific quantities. In the case of a shock wave $\oo^{\m\n}=0$.  In order to examine the asymptotic falloff in a linearized theory we perturb the asymptotic values of $u,T$ with \eno{SECANSATZ} and, in addition, one component of the $\pi^{\m\n}$ tensor with
\eqn{PZZPERT}{
\pi^{00}(x) = \d \pi^{00} e^{iqx}\;.
}  
The resulting three by three system has a solution only if $q$ is given by 
\eqn{GOTQIS}{
\frac{iq}{\pi T} =  \frac{3v^2-1}{v}\frac{\sqrt{1-v^2}}{\bar{\h}+\bar{\t}_{\pi}\left(1-3v^2\right)}\;.
} 
Notice that there is a pole, which for the $\mathcal{N}=4$ values (\ref{SCFTIS}) is located in $v\sim .92$.
On the other hand, 
the fully microscopic calculation based on the gravity dual (and described in section \eno{STRONGSH}) shows
that $q$ remains finite at all $v < 1$. This may lead one to choose 
$\bar{\h},\bar{\t}_{\pi}$ in such a way the pole in \eno{GOTQIS} is located at $v=1$.  However, even with this choice the Israel-Stewart theory fails to capture the asymptotic behavior of $q$ at $v\rightarrow 1$.  
Equation \eno{GOTQIS} predicts that $q$ increases linearly with $\g=\frac{1}{\sqrt{1-v^2}}$, whereas the linearized gravity analysis of section \eno{STRONGSH} predicts  $q \propto \g^{1/2}$. 
 
\subsection{Effective Hydrodynamics}\label{EFFHYDRO}

Effective hydrodynamics is the approach where one attempts to model the effect of higher-order 
terms in the gradient
expansion of $T^{\mu\nu}$ with terms that are high in derivatives but linear in velocity.
Such an approach has been taken up in Ref.~\cite{Lublinsky:2009kv} and
may seem an ideal way to encode results of a linearized theory. Effective hydrodynamics for a
conformal theory in flat space can be summarized by writing the stress tensor as 
\eqn{DEFTMNEFF}{
T^{\m\n}=p \h^{\m\n} +\left(\r+p\right)u^{\m}u^{\n} +\Pi^{\m\n}\;,
}
\eqn{DEFPEFF}{
\Pi^{\m\n} = -2 \int dt' \int d^{3} x' D(x-x',t-t')\s^{\m\n}(x',t')\;,   
}  
\eqn{DEFDEFF}{
D(x,t) = \int d\oo d^3 k e^{-i\oo +i kx} \h (\oo,k^2)\;.
}
The effective viscosity $\h (\oo,k^2)$ is taken to be a function of $\oo,k^2$ such that it correctly 
reproduces the location of the poles of the scalar, shear and sound modes up to the desired order.
For the $\mathcal{N}=4$ plasma, it
has been calculated up to the fifth order in \cite{Lublinsky:2009kv} and found to be 
\eqn{GOTEFFETA}{
\h = \h_{0} \left(1+i \h_{0,1}\oo +\h_{0,2} \oo^2 +i \h_{2,1} \oo k^2 +i \h_{0,3} \oo^3+ \h_{4,0} k^4 +\h_{2,2}\oo^2k^2 + \h_{0,4}\oo^4 +\cdots \right)\;,
}   
\eqn{GOTETACOEF}{
\h_{0}=\frac{\r+p}{2}\;,\quad \h_{0,1}=2-\ln 2\;,\quad  \h_{2,0}=-\frac{1}{2}\;, \cr 
\h_{0,2} \approx -1.379 \;,\quad \h_{2,1} \approx -2.275\;,\quad \h_{0,3} \approx -0.082\;, \cr
\h_{4,0}\approx 0.565\;,\quad  \h_{0,4}\approx 2.9\;, \quad \h_{2,2} \approx 1.1\;, 
}
where we have not given the uncertainties of each coefficient. One can attempt to resum this fifth-order 
expression into a rational expression with one or two poles \cite{Lublinsky:2009kv}. 

To compute the asymptotic tails of the shock in effective hydrodynamics, we consider again perturbations of 
the type \eno{SECANSATZ} in the rest frame of the shock. The result is given by 
\eqn{GOTQEFFH}{
\frac{iq}{\pi T} =  \frac{3v^2-1}{v}\sqrt{1-v^2}\frac{\h_{0}}{\h(-\frac{v q}{\sqrt{1-v^2}},\frac{q^2}{1-v^2})}\;,
} 
where $v$ is the speed of the fluid relative to the shock. The actual curve 
$q=q(v)$ is determined by linearized gravity and can only be found numerically;
the result is shown in Fig. \eno{Qcomp}.  A simple ansatz for the effective viscosity does not reproduce
this curve very well. For example an effective viscosity 
with one or two poles will always give $q\rightarrow 0$ as $v\rightarrow 1$, in contrast to the behavior 
following from linearized gravity. On the other hand, as seen in Fig. \eno{Qcomp}, the expansion of 
Ref.~\cite{Lublinsky:2009kv} gives a reasonable approximation for $q$ on the subsonic side 
of the shock. 

We can take the idea of effective hydrodynamics one step further by simply encoding our numerical
curve into an expression for $T^{\mu\nu}$, so that agreement with linearized gravity is perfect by
construction. In general, any hydrodynamic approximation amounts to reconstructing $T^{\mu\nu}$ from
its timelike eigenvector $u^{\m}$ (cf. Eq.~\eno{DEFTU}). Consider the fluid at rest with a sound wave of small 
amplitude $\varepsilon$ propagating along $x$ with momentum $k$ and (in general complex) frequency 
$\omega(k)$. From conservation of $T^{\m\n}$ 
and the traceless condition we find that the Fourier components of $\d T^{\m\n}$ are given by
\beq
\d T^{tt} = \varepsilon, \ \ \d T^{tx} = \frac{\omega}{k} \varepsilon, \ \ \d T^{xx}=\frac{\omega^2}{k^2} \varepsilon, \ \ 
\d T^{yy}=\d T^{zz} =\frac{k^2-\omega^2}{2k^2} \varepsilon \;.
\eeq
It is now a simple matter to compute the timelike eigenvector of $T^{\m\n}$, identify $\d u=\frac{\omega}{4q T_0^4}\varepsilon$ as the $x$ component of the
four velocity and write $\d T^{\mu\nu}$ as 
\beq
\d T^{\mu\nu} = \d T_{(0)}^{\m\n} + \d T_{(1)}^{\m\n}
\eeq
where $\d T_{(0)}^{\m\n}$ is the variation of the ideal fluid energy momentum tensor $T_{(0)}^{\mu\nu}= (\pi T_{0})^4 \left(\eta^{\m\n}+4 u^\m u^\n\right)$  and
$\d T_{(1)}^{\m\n}$ is an extra contribution given by
\beqa
\d T_{(1)}^{xx} &=&\frac{4}{3k\omega}\left(3\omega^2-k^2\right) T_0^4 \d u,  \\
\d T_{(1)}^{yy}&=&\d T_{(1)}^{zz} = -\frac{2}{3\omega k} \left(3\omega^2-k^2\right) T_0^4 \d u .
\eeqa
In these expressions $\omega$ should be understood as a function of $k$ obtained by solving numerically the gravity equations for the sound wave. Notice that 
the function $\frac{3\omega^2-k^2}{\omega k}$ is regular for $k\rightarrow 0$, so $T_{(1)}^{\m\n}$ is a well defined function of $\d u$. It is 
non-local since it involves an infinite number of derivatives. Nevertheless, in the linear approximation we can work with this energy momentum 
tensor that reproduces exactly the sound pole and the asymptotic behavior of the shock wave far from the shock. 
Later, in the numerical section we give an approximate result for the function $\omega(k)$ that could be used, if so desired, to further simplify and 
approximate $\d T_{(1)}^{\m\n}$. Such effective hydrodynamics is still not sufficient to describe the 
center of the shock, where the linearized approximation is not applicable, but at least 
it summarizes all the information we were
able to extract from gravity without attempting to find the full numerical solution to the Einstein equations 
in the bulk. 

\begin{figure}
 \begin{center}
\includegraphics[width=12cm]{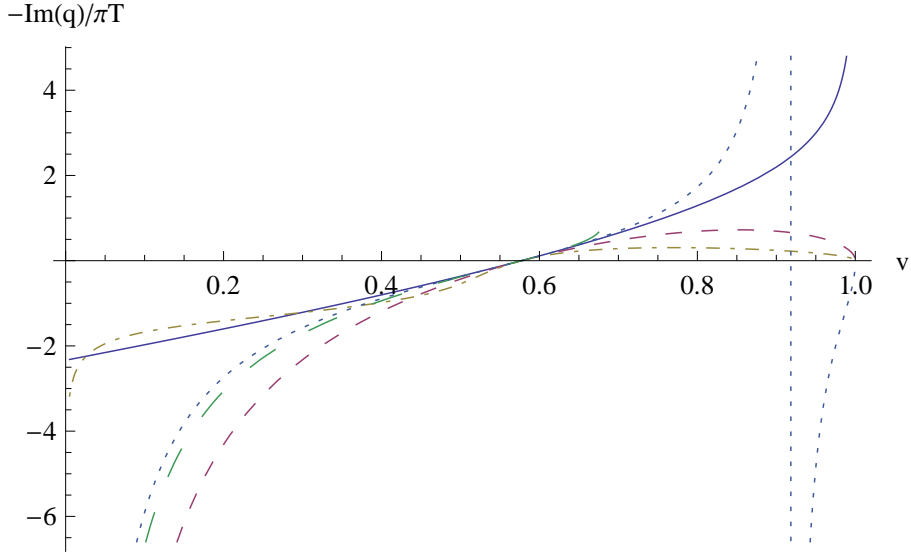}
 \end{center}
 \caption[]{The solid (blue) line represents the numerical results discussed in section \eno{STRONGSH}.  The (purple) dashed line is the first order hydrodynamics result and the (green) dashed line is the second order result.  The (blue) dotted line is the Israel-Stewart result and the dotted-dashed line is the result for the effective hydrodynamics theory of section \eno{EFFHYDRO}. None of the theories captures the $v\rightarrow 1$ asymptotics of the numerical result.}
\label{Qcomp}
 \end{figure} 

\section{Shock waves in the gravity-hydrodynamics correspondence}\label{SHOCKGR}

 In a strong shock wave, the region near the shock is beyond the hydrodynamic approximation and can only be described as a jump in the hydrodynamic
quantities. In the case of the strongly coupled ${\cal N}=4$ plasma, hydrodynamics ceases to be valid at distances shorter than the inverse temperature. 
However, at those distances the bulk description in terms of gravity does not break down suggesting that gravity can resolve
the shock waves and provide a smooth description for them. However the velocity and temperature are not well defined in the region of the 
shock, so the best characterization is in terms of the energy density, namely $T^{tt}(x)$. In this section we first study such function in the case of 
weak or hydrodynamic shocks. In that case we can reproduce the results of the previous section and obtain the dual metric 
(within the hydrodynamic approximation). Afterwards we consider strong shocks and, by using linearized gravity,
obtain the width of the 
shock, as determined by the exponential tails on both sides. This is detailed information that can only be obtained from a microscopic theory of the system. In principle, one would like to go further 
and obtain $T^{tt}(x)$ by solving the Einstein equations in the bulk numerically, but such a calculation is beyond the scope 
of this paper.

\subsection{Weak shocks: an explicit solution in the fluid-gravity correspondence}\label{FLUIDGR}

 The four conservation equations $\partial_\mu T^{\mu\nu}=0$ are not enough to determine the nine independent components 
of $T^{\mu\nu}$ in the boundary theory. Hydrodynamics amounts to a restriction on $T^{\mu\nu}$ by providing an
expression for it in terms of four variables, 
the velocity $v_{i=1,2,3}$ and temperature $T$, which are then determined from the conservation equations. 
The expression for $T^{\mu\nu}$ is given as a derivative 
expansion, and its precise form can only be determined from a microscopic theory of the system.  

 From the dual gravitational point of view, the energy momentum tensor sets the boundary conditions for an asymptotically AdS metric and the conservation
equations $\partial_\mu T^{\mu\nu}=0$ are necessary consistency conditions for the existence of a solution to the Einstein equations with
such boundary data.  As recently explained by Bhattacharyya et al. \cite{Bhattacharyya:2008jc} those metrics generically have naked 
singularities. The condition of the metric being non-singular imposes a restriction to $T^{\mu\nu}$, which is a counterpart of the restriction seen in the
hydrodynamic construction. In fact, when this analysis is done in a derivative 
expansion, as shown in \cite{Bhattacharyya:2008jc}, it provides a microscopic derivation of the hydrodynamic equations for the strongly coupled ${\cal N}=4$ plasma.  

 In this section we use the BHMR construction \cite{Bhattacharyya:2008jc} to obtain the metric dual to the shock waves in the hydrodynamic regime up to terms
which are third order in the derivative expansion.  

 Let us start by summarizing the procedure as adapted to our particular problem. The starting point is the boosted black brane in 
Eddington-Filkenstein coordinates:
\beq
ds_{(0)}^2 
 = -2u_{\mu} dx^{\mu}dr + \frac{\left(\pi T(x)\right)^4}{r^2} u_{\mu}u_{\nu}dx^{\mu}dx^{\nu} + r^2 \eta_{\mu\nu}dx^{\mu} dx^{\nu}\;,
\eeq
where 
\beq
u^{\mu} = (u^0(x),u(x),0,0), \ \ \ \ \ u^0(x) = \sqrt{1+u(x)^2}\;.
\eeq
 Since $u(x)$ and $T(x)$ are not constant this metric does not solve the Einstein equations
\beq
 \Gg_{MN} = R_{MN}+4 g_{MN}=0\;,
\eeq
 where $M,N=r,t,x,y,z$ denote 5-dimensional indices (whereas $\mu,\nu=t,x,y,z$ denote four-dimensional indices). 
 To find a solution we introduce a formal parameter $\epsilon$ and expand
\beqa
 u^{\mu} &=& u_{\mu}^{(0)} + \epsilon u_{\mu}^{(1)} + \epsilon^2 u_{\mu}^{(2)}+ \epsilon^3 u_{\mu}^{(3)} + {\cal O}(\epsilon^4)\;,  \label{uexp}\\
 T(x) &=& T_{(0)}(x) + \epsilon T_{(1)}(x) + \epsilon^2 T_{(2)}(x) + \epsilon^3 T_{(3)}(x)  + {\cal O}(\epsilon^4)\;.
\eeqa
  At the same time, the metric is corrected by adding an expansion 
\beq
 ds^2= ds_{(0)}^2 + \epsilon ds_{(1)}^2 + \epsilon^2 ds_{(2)}^2  + {\cal O}(\epsilon^3)\;.
\eeq
 The tensor $\Gg_{MN}$ is expanded in powers of $\epsilon$ with the rule that each $x$-derivative counts as one extra power of $\epsilon$. The equations 
are solved order by order. For that purpose it is convenient to introduce the vector
\beq
\tilde{u}^{\mu} = (u(x), u^0(x),0,0)\;,
\eeq
orthogonal to $u^\mu$. In this way the metric corrections can be parametrized as
\beqa\label{GOTBHMRSOL}
 ds^2_{(n)} &=& (s_1^{(n)} +s_2^{(n)}) u_{\mu}u_{\nu}dx^{\mu}dx^{\nu} + s_2^{(n)} \eta_{\mu\nu}dx^{\mu} dx^{\nu} + 2 s_3^{(n)} u_{\mu} dx^{\mu}dr
              +  s_V^{(n)}  \tilde{u}_{\mu}u_{\nu}dx^{\mu}dx^{\nu} \nonumber \\ 
             && + s_T^{(n)} \left( \tilde{u}_{\mu}\tilde{u}_{\nu}dx^{\mu}dx^{\nu} -\half(dy^2+dz^2)\right)\;,
\eeqa
where the functions $s_{1,2,3}^{(n)}, s_V, s_T$ describe scalar, vector and tensor perturbations classified according to the local $SO(3)$ group that 
leaves $u^{\mu}$ invariant. Notice that each $ds^2_{(n)}$ should in turn be expanded using (\ref{uexp}). 
 Following \cite{Bhattacharyya:2008jc} we make the gauge choice $g^{(n)}_{rr}=0$, $g^{(n)}_{\mu r}\sim u_{\mu}$ and $g_{(0)}^{\mu\nu} g^{(n)}_{\nu\mu}=0$ for all $n>0$.
In that case it is convenient to parametrize the fluctuations as
\beqa
s_1^{(n)} = &\frac{1}{r^2} k^{(n)}(x,r), \ s_2^{(n)} = r^2 h^{(n)}(x,r), \ s_3^{(n)} = \frac{3}{2} h^{(n)}(x,r), \cr
 & \ s_V=\frac{1}{r^2} j^{(n)}(x,r), \ s_T^{(n)} = r^2 \alpha^{(n)}(x,r)\;,
\eeqa
 In order to solve the equations order by order it is convenient to decompose $\Gg_{MN}$ into its scalar,
vector and tensor parts which decouples the equations. The procedure is in principle straightforward, and
we proceed to describe the results. 
\begin{description}
\item[\underline{Order 1}.]
 The first equations that we obtain are $u_{0}'=0$, $T_{(0)}'=0$ implying that 
\beq
u_{(0)}(x) = u_{(0)}, \ \ \ T_{(0)}(x) = T_{(0)}\;,
\eeq
namely they are constant functions. In that case the zero order metric is an exact solution and there is no first order correction to the metric:
\beq
 h^{(1)}= k^{(1)}=j^{(1)}=\alpha^{(1)}=0\;.
\eeq
The temperature however is corrected to
\beq
T^{(1)} = -\frac{\sqrt{2}}{3} T_{(0)} u_{(1)} \;.
\eeq
\item[\underline{Order 2}.]
The first equation we find is
\beq
 u_{(1)}' (2u_{(0)}^2-1)=0\;.
\eeq
We can take $u_{(1)}$ constant which leads us  to a trivial solution, or otherwise we require $u_{(0)}=\frac{1}{\sqrt{2}}$ implying that the zero order 
solution describes a fluid moving at the speed of sound, which is the appropriate starting point to describe shocks in the hydrodynamic approximation. 
The other equations give:
\beq
 h^{(2)}=0, \ \ k^{(2)}=\frac{2}{3} r^3 u_{(1)}' , \ \ j^{(2)}=-\frac{2}{\sqrt{3}} r^3 u_{(1)}', 
 \ \ \alpha^{(2)}= \frac{u_{(1)}'}{3\pi T_{(0)}} F_1\left(\frac{r}{\pi T_{(0)}}\right)\;,
\eeq
where
\beq
F_1(y) = \ln\left(\frac{(1+y^2)(1+y)^2}{y^4}\right) - 2\arctan y +\pi\;.
\eeq
Notice that at this order $u_{(1)}$ is undetermined. This is a particular property of our solution that requires us to go to higher orders to obtain 
the metric. 
\item[\underline{Order 3}.]
The first equation we obtain is
\beq
 u_{(1)}'' = \frac{8}{3} \pi T_{(0)} u_{(1)} u_{(1)}' \;, \label{u1eq}
\eeq
which allows us to solve for $u_{(1)}$ as we do further below. The components of the metric are corrected by
\eqn{SECCORR}{
 h^{(3)}&=0\;, \quad k^{(3)}=\frac{2}{3} r^3 u_{(2)}' - \frac{\sqrt{2}}{3} r^2 u_{(1)}'' , \cr
 j^{(3)}&=-\frac{2}{\sqrt{3}} r^3 u_{(2)}' + \frac{4\sqrt{2}}{3\sqrt{3}} u_{(1)}u_{(1)}'(\pi T_{(0)})^3  F_2\left(\frac{r}{\pi T_{(0)}}\right), \cr \\
 \alpha^{(3)}&= \frac{u_{(2)}'}{3\pi T_{(0)}} F_1\left(\frac{r}{\pi T_{(0)}}\right) - \frac{4\sqrt{2}}{9\pi T_{(0)}} 
  u_{(1)}u_{(1)}' F_3\left(\frac{r}{\pi T_{(0)}}\right)\;.
}
with
\beqa
F_2(y) &=& \half(y^4-1)\left(2\arctan y+ \ln\left(\frac{1+y^2}{(1+y)^2}\right)\right)-\half\pi y^4+y^3+y^2-\frac{25}{12}\;, \\
F'_3(y) &=& \frac{1}{y^5-y}\left\{2(1-y^3)\left(\arctan y-\frac{\pi}{2}-\ln(1+y)\right)+(1+y^3)\ln(1+y^2)\right.  \nonumber\\
        && \left. -4y^3\ln y  +7-2\ln2-\frac{2(1+y)}{1+y^2}-\frac{2}{1+y}-4y^2 \right\}\;.
\eeqa 
 We give the derivative of $F_3$ since it is the function that enters in subsequent calculations. It can be integrated explicitly 
in terms of dilogarithms but the expression is not very illuminating.  
The temperature is given by
\beq
\pi T_{(2)} = -\frac{\sqrt{2}}{3} \pi T_{(0)} u_{(2)} +\frac{1}{3}u_{(1)}'\;.
\eeq
We want to write the metric up to order $\epsilon^2$ which requires computing $u_{(2)}$. Since it is undetermined at this order we continue the expansion.
\item[\underline{Order 4}.]
 At this order we only look for the equation determining $u_{(2)}$. However we need to include and keep track of the terms $u_{(3)}$, $T_{(3)}$ , $h_{(4)}$ etc. 
to be sure that they do not appear in such equation. What we get is
\beq
 u_{(2)}'' - \frac{8}{3}\pi  T_{(0)} (u_{(1)} u_{(2)})'= \frac{\sqrt{2}}{3} (7-2\ln2)u_{(1)}'{}^2-\frac{4\sqrt{2}}{9}\pi T_{(0)} u_{(1)}^2u_{(1)}'(1+4\ln 2)\;.
\eeq
\end{description}
The last equation, together with (\ref{u1eq}), can be easily solved to obtain, at this order
\beqa
 u &=& u_{(0)} + u_{(1)} + u_{(2)}\;, \\ \nonumber\\
 u_{(0)} &=& \frac{1}{\sqrt{2}}\;,  \\
 u_{(1)} &=& - u_{\infty} \tanh\xi\;, \\
 u_{(2)} &=& \frac{u_{\infty}^2}{6}\left[4\sqrt{2}(1-\ln 2) \frac{\ln\cosh\xi}{\cosh^2\xi}
             +5\sqrt{2}\left(\tanh^2\xi+\tanh\xi+\frac{\xi}{\cosh^2\xi}\right)\right] \;,
\label{uexp2}
\eeqa
where
\beq
\xi = \frac{4\pi T_{(0)} u_{\infty}}{3}\, x\;,
\eeq
 and we set the formal parameter $\epsilon=1$. It is interesting to note that the equation for $u_{(1)}$ determines that the fluid reaches the shock wave 
supersonically and leaves subsonically. In other words, we are not free to exchange the subsonic and
supersonic sides of the shock. The reason is that are choosing gravity solutions which are regular in the
infalling Eddington-Filkenstein 
coordinates as appropriate for a black hole. 
 The constant $u_{\infty}$ is arbitrary and determines the amplitude of the shock, namely, the asymptotic value of the velocity. For consistency of the 
approximation we require $u_{\infty}\ll \frac{1}{\sqrt{2}}$. In fact, $u_{\infty}$ plays the role of the small parameter, as can be seen from the fact that the 
velocity depends on $x$ through $\xi$ and so each $x$ derivative brings in an extra power of $u_\infty$. The behavior at infinity
is given by
\beqa
u(x\rightarrow\pm\infty)&\simeq& \frac{1}{\sqrt{2}} \mp u_{\infty} 
                                 \pm 2u_{\infty} e^{\mp2\xi}\left(1+\frac{1}{3}u_{\infty} (4\sqrt{2}-4\ln2\pm5\sqrt{2})\xi\right) \\
                        &\simeq&  \frac{1}{\sqrt{2}} \mp u_{\infty} \pm 2u_{\infty} e^{\mp2\xi+\frac{1}{3}u_{\infty} (4\sqrt{2}-4\ln2\pm5\sqrt{2})\xi} \;.
\label{qgrav}
\eeqa
We write the correction in the exponential form for an easier comparison with the hydrodynamic result.
For the temperature we have
\beq
T(x\rightarrow\pm\infty) = T_{(0)} \pm \frac{\sqrt{2}}{3} T_{(0)} u_\infty\;.
\eeq
For the velocity we have
\beqa
v(x\rightarrow\infty) &\simeq& \frac{1}{\sqrt{3}} +\delta v_+ = \frac{1}{\sqrt{3}} 
                                        -\frac{2\sqrt{2}}{3\sqrt{3}} u_\infty + \frac{14}{9\sqrt{3}} u_\infty^2\;, \\
v(x\rightarrow-\infty) &\simeq& \frac{1}{\sqrt{3}} +\delta v_- = \frac{1}{\sqrt{3}} 
                        +\frac{2\sqrt{2}}{3\sqrt{3}} u_\infty - \frac{2}{3\sqrt{3}} u_\infty^2 \;.
\eeqa
The first check is that the condition (\ref{GTOV2})
\beq
v(-\infty) v(+\infty) = \frac{1}{3} +{\cal O}(u_\infty^3)\;,
\eeq
is satisfied to the considered order. From eq.~(\ref{qgrav}), the exponential tail is given by
\beq
 v(x\rightarrow \pm\infty) = \frac{1}{\sqrt{3}} + \delta v_\pm + c_{\pm} e^{iq_{\pm} x}\;,
\eeq
for some constants $c_{\pm}$ and the penetration depth determined by
\beq
\frac{iq_{\pm}}{\pi T_{\pm}} = 2\sqrt{6} \delta v_\pm +6\sqrt{2} \delta v_{\pm}^2 (1-\ln 2)\;,
\eeq  
in complete agreement with the hydrodynamic calculation. This is not surprising since the hydrodynamic 
equations arise from gravity.
The calculation in this section, however, allows us to compute in addition the dual metric, which 
to this order is given by
\beqa
ds^2 &=& -2u_\mu dx^\mu dr + \frac{1}{r^2} ((\pi T)^4+k) u_\mu u_\nu dx^\mu dx^\nu + r^2 \eta_{\mu\nu} dx^\mu dx^\nu +
         \frac{1}{r^2} j \tilde{u}_\mu u_\nu dx^\mu dx^\nu \\
     && + \alpha r^2(\tilde{u}_\mu\tilde{u}_\nu dx^\mu dx^\nu -\half(dy^2+dz^2)) \;,
\eeqa
where $u^\mu=(\sqrt{1+u^2},u,0,0)$, $\tilde{u}^\mu=(u,\sqrt{1+u^2},0,0)$ and $u$ should be expanded as  
$u=u_{(0)}+u_{(1)}+u_{(2)}$ using the $u_{(n)}$ computed in (\ref{uexp2}). The temperature is also expanded as
\beq
T = T_{(0)} -\frac{\sqrt{2}}{3} T_{(0)} u_{(1)} -\frac{\sqrt{2}}{3} T_{(0)} u_{(2)} +\frac{1}{3\pi}u_{(1)}'\;,
\eeq
whereas the other functions entering the metric are given by
\beqa
 k &=& \frac{2}{3} r^3 (u_{(1)}' +  u_{(2)}') - \frac{\sqrt{2}}{3} r^2 u_{(1)}'' \;, \\
 j &=& -\frac{2}{\sqrt{3}}\, r^3 (u_{(1)}' + u_{(2)}') 
       + \frac{4\sqrt{2}}{3\sqrt{3}} u_{(1)}u_{(1)}' (\pi T_{(0)})^3  F_2\!\left(\frac{r}{\pi T_{(0)}}\right), \\
 \alpha &=& \frac{u_{(1)}'+u_{(2)}'}{3\pi T_{(0)}} F_1\!\left(\frac{r}{\pi T_{(0)}}\right)
            - \frac{4\sqrt{2}}{9\pi T_{(0)}} u_{(1)}u_{(1)}' F_3\!\left(\frac{r}{\pi T_{(0)}}\right)\;,
\eeqa
with the $F_{1,2,3}$ as defined above. The expansion parameter is the strength of the shock as determined by the constant $u_\infty$ 
appearing in $u_{(n)}$.
\section{Strong shocks: the linearized gravity approximation}\label{STRONGSH}
Strong shocks are characterized by large gradients of velocity and temperature
and cannot be studied within hydrodynamics: we can say that hydrodynamics does
not resolve their profiles. However, the gauge-gravity correspondence is not 
limited to small gradients, and so the gravity side of it
should contain information about strong shocks as well. We now discuss how that
information can be extracted.

In principle, we expect that there are exact solutions to the 5-dimensional 
Einstein equations, and those solutions describe strong shocks exactly. 
This belief is based on the observation that the asymptotic values of $u_\mu$
and $T$ on the far left and far right of the shock have vanishing gradients
and are therefore well reproduced even by the ideal hydro (see Sec.~\ref{SHHYDRO}).
On the gravity side, to each of these asymptotics, there corresponds a
5-dimensional AdS black brane, suitably boosted and with a suitable
value of the temperature. Then, there must be a 5d solution describing a stationary wave
that smoothly interpolates between these two regions---the gravity dual of a strong shock.

The exact solution (assuming it exists) described in the preceding paragraph 
would tell us all there is to know about a strong shock, in particular, the
profile of the average energy density, $\epsilon(x)$. 
So far, however, we have not been able to find any such solution explicitly.
In this section, we provide partial information about the profile of 
a strong shock, obtained by looking at linearized gravity on the backgrounds 
corresponding to each of the two asymptotic regions ($x\to \pm \infty$). 

\subsection{Equations of linearized gravity}
In linearized gravity, one writes the metric in the form 
$g_{\mu\nu}= g^{(0)}_{\mu\nu} + h_{\mu\nu}$, where $g^{(0)}_{\mu\nu}$ is the metric
of the AdS black brane, and $h_{\mu\nu}$ is a perturbation, and works to the 
first order in the perturbation. Solutions to the linearized Einstein equations are
known as quasinormal modes. 
In this paper, we consider solutions that depend only on the coordinate ($x$)
along the direction in which the shock propagates and, possibly, time.  In this section we adopt the convention $\pi T=1$, the temperature dependence can be recovered by multiplying the modes by $\pi T$.  Due to the translational invariance along the brane
directions, we can search for these solutions in the form
\[
h_{\mu\nu}(t,x,r) = r^2 H_{\mu\nu}(r) e^{-i\omega t + i q x} \, .
\]
We adopt the convention that $\omega$ and $q$ refer to the boosted frame, moving
at the speed of the shock, and their primed counterparts, $\omega'$ and $q'$, to
the unboosted frame, connected with the fluid. Note that there are actually two
such unboosted frames (the fluid on the two sides of the shock moves
with different velocities), but in the linearized approximation the two sides
are disconnected and can be considered separately.

Linearized gravity has by now become a familiar tool in studies of the kinetics 
of the strongly coupled ${\cal N}=4$ plasma but in a setting that is typically
different from ours. In many cases (as, for example, in the 
computation of the viscosity \cite{Policastro:2001yc}) one considers relaxation
of an initial perturbation. Then, one picks a real wavenumber $q$ and looks for the 
corresponding (complex) frequencies. Here, in contrast, we are interested in propagation
of a boundary disturbance, that is, in how
far a perturbation with a given frequency extends into the plasma on either side 
of the shock.
For this, we pick a real $\omega$ and look for the corresponding (complex) $q$.
Specifically, we will be interested in perturbations with $\omega=0$, as we expect
these to describe the behavior of the average energy density of a shock wave 
sufficiently far away from it.\footnote{To be sure, it is not obvious a priori
that the shock wave profile is static: there could be instabilities in the nonlinear
central region that cause oscillating behavior. In the linearized theory, a possible
signal of such an instability would be the absence of a physically acceptable 
static solution (due, for instance, to a singularity in the equation). 
We have not found any such signals in our calculations.}

A classification of the quasinormal modes of an (unboosted) AdS black brane has been
given in \cite{Kovtun:2005ev}; a comprehensive recent review of quasinormal modes is \cite{Berti:2009kk} . According to that classification, metric fluctuations group into 
several channels, corresponding to different gauge-invariant combinations of 
the components of $H_{\mu\nu}$. Here, we are interested in the sound channel. 
The corresponding quasinormal modes satisfy the equation \cite{Kovtun:2005ev}
\be
Z'' + P(u) Z' + Q(u) Z = 0 \, ,
\label{QN2}
\ee
where primes denote derivatives with respect to $u = r_0^2 /r^2$, 
\ba
P(u) & = & - \frac{3 {\omega'}^2 (1+u^2) + {q'}^2 (2u^2 - 3 u^4 - 3)}
{u f(u) [3 {\omega'}^2 + {q'}^2 (u^2 - 3) ]}  \, , \\ 
Q(u) & = & - \frac{4 {q'}^2 u^2}{f(u) [3 {\omega'}^2 + {q'}^2 (u^2 - 3) ] }
+ \frac{ 3 {\omega'}^4 + {q'}^4 (u^4 - 4u^2 + 3)  + {\omega'}^2 {q'}^2 (4 u^2 - 6)}
{4 u f^2(u) [3 {\omega'}^2 + {q'}^2 (u^2 - 3) ]}
\, ,
\ea
and $f(u) = 1-u^2$. Remember that in these expressions,
$\omega'$ and $q'$ are in units\footnote{And so are twice as large as their 
counterparts in \cite{Kovtun:2005ev}; hence an extra overall $\frac{1}{4}$ in the second term
in $Q$.} of $\pi T$ and refer to the unboosted frame. They are
related to the frequency and wavenumber in the boosted frame by the Lorentz 
transformation
\begin{eqnarray}
\omega' & = & \omega \cosh \beta - q \sinh \beta \, , \label{LT1} \\
q' & = & - \omega \sinh \beta + q \cosh \beta \, , \label{LT2}
\end{eqnarray}
where $\tanh \beta = v$, the speed of the shock. For static perturbations, we set
$\omega = 0$ and substitute the resulting expressions for $\omega'$ and $q'$ into
\eq{QN2}, to obtain:
\begin{eqnarray}
P(u) & = & \frac{3 + 3 u^2 - 5 \g^2 u^2 + 3 \g^2 u^4}{u f(u) (\g^2 u^2 - 3)} \, , 
\label{P} \\
Q(u) & = & - \frac{4 \g^2 u^2}{f(u) (\g^2 u^2 -3)} + q^2 \frac{\g^2 u^2 - 1}{4 u f^2(u)}
\label{Q} \, ,
\end{eqnarray}
where we have used the shorthand $\g = \cosh \beta$.

The same expressions can be obtained by starting directly in the 
boosted frame. In this case, the unperturbed metric is
\[
ds^2_0 = r^2 \eta_{\mu\nu} dx^\mu dx^\nu 
+ \frac{r_0^4}{r^2} (dt \cosh\beta - dx \sinh \beta)^2 
+ \frac{dr^2}{r^2 (1 - r_0^4/r^4)} \, ,
\]
and the perturbation reads
\[
ds^2_1 = r^2 \left[ H_{00} dt^2 + H_{11} dx^2 + 2 H_{01} dt dx + H (dy^2 + dz^2)
\right] e^{iqx}
\]
(all $H_{\mu\nu}$ are functions of $r$ only).
The relevant gauge-invariant combination (at $\omega = 0$) is 
\eqn{DEFINZ}{
Z(r) = H_{00}(r) + \left( 1 + \frac{r_0^2}{r^4} \g^2 \right) H(r)\;,
}
and satisfies \eq{QN2} with the coefficient functions given by \eqs{P}{Q}.

For computation of properties of the plasma via the gauge-gravity correspondence, 
we only need to consider \eq{QN2} in the region outside the horizon,
$r_0 < r < \infty$, which we will refer to as the physical region. In terms of
the variable $u$, it corresponds to $0 < u < 1$.
A noteworthy property of the coefficient functions (\ref{P}) and (\ref{Q}) is 
that, at sufficiently large boost velocities,
\be
\cosh \beta > \cosh \beta_{\rm cr} = \sqrt{3} \, ,
\label{betacr}
\ee
$P$ and $Q$ both have poles inside the physical region, at 
\eqn{FUNNY}{
u = u_1 \equiv \frac{\sqrt{3}}{\cosh\beta} \, .
}
From the outset, we might have anticipated that we would need to impose boundary
conditions at the boundaries  of the physical region, $u\to 0$ and $u\to 1$, but 
not at any interior point. We therefore need to
explore the nature of the singularity at $u=u_1$ in more detail.

Let us search for solutions near $u=u_1$ in the form $Z\sim (u - u_1)^s$. 
For the exponent $s$, we find two roots,
\eqn{GOTINDICES}{
s= 0 \mbox{ or }  3.
}
Fuchs's theorem \cite{Arfken} guarantees that the larger root correspond to a regular solution,
expandable in powers of $w = u - u_1$ as follows: $Z = w^3 + O(w^4)$. As for the 
solution corresponding to the smaller root, in general, we expect it to have the
form
\be
Z(u) = c_0 + c_1 w + c_2 w^2 + c_3 w^3 + c_3' w^3 \ln w + \ldots 
\label{Zexp}
\ee
The recursion equation for the coefficients $c_n$, which is obtained by substituting
\eq{Zexp} in \eq{QN2}, degenerates at the order (and only at the order) at which
the second solution appears, in our case the order $w^3$. 
Whether or nor the logarithmic term in (\ref{Zexp}) is nonzero
then depends on the precise values of the coefficients of all the terms up to $O(w)$ order in the expansions of
$P(u)$ and $Q(u)$. As it turns out, there is a curious cancellation among these
terms, such that $c_3' =0$ for all
values of $\g$ and $q$. We conclude that both solutions are regular at $u=u_1$, and
a boundary condition there is not required.

\subsection{Boundary conditions. Irreversibility} \label{BCIRR}
The choice of boundary conditions for quasinormal modes that is suitable for
applications of the gauge-gravity duality to kinetic theory
has been discussed in the literature
(see, for example, Ref.~\cite{Policastro:2001yc}), and we do not deviate from 
it here. Our computation, however, requires an analytical continuation of 
these boundary conditions, which is the subject of this subsection.

At $u\to 0$ (the boundary of the AdS space) we use the standard
\eqn{bc0}{
Z(0) = 0 \, .
}
At $u\to 1$ (the near-horizon region), we first consider real $\omega'$ and pick, 
as usual, the wave infalling with respect to the black brane:
\eqn{infall}{
Z(u\to 1) \sim (1-u)^{-i \omega'/4} \, .
}
For the present problem, having to do with propagation of a perturbation in space,
rather than in time, we need to
analytically continue this expression to complex $\omega'$ given by
the Lorentz transformation (\ref{LT1}) ($\omega'$ is complex because
so is $q$). In particular, for the static case ($\omega=0$), we
have $\omega' = -q \sinh\beta$ and thus
\eqn{bc1}{
Z(u\to 1) \sim (1-u)^{\frac{1}{4} i q \sinh\beta} \, .
} 
This choice is equivalent to choosing the solution that is regular in infalling Eddington-Filkenstein coordinates
as done in section \ref{FLUIDGR}. There is an exceptional case $\beta =\beta_{\rm cr}$, 
the critical value given by \eq{betacr}. For this value of $\beta$, 
the analytical 
continuation to $\omega=0$ causes confluence of the singularities
at $u=1$ and $u=u_1$, which modifies the asymptotic behavior near
$u=1$. We consider this case separately later in this subsection.

Recall that $Z(u)$ corresponds, via the gauge-gravity duality, to a perturbation
in the plasma that depends on $x$ as $\exp(iqx)$. On physical grounds,
we expect that perturbations corresponding to the tails of a shock at 
$x\to\pm\infty$ decay away from the shock. This means that we must pick
$q$ with a positive (negative) $\imq$ for the fluid at positive (negative)
$x$. Recall also that positive (negative) $x$ correspond to the subsonic 
(supersonic) side of the shock. Thus, according to \eq{bc1}, $Z(u)$ is regular
at the horizon on the supersonic side but singular on the subsonic one.

Formulating a boundary problem for the regular case presents no difficulty:
the second solution to \eq{QN2} diverges at $u\to 1$, and the boundary condition
(\ref{bc1}) selects the one that does not. The singular case (${\rm Im}q>0$) is 
a bit trickier: we wish to retain the divergent solution and reject the convergent one.
To achieve that, we peel off the singular part, as follows:
\eqn{peel}{
Z(u) = (1-u)^s \Psi(u) \, ,
}
where
\eqn{sdef}{
s = - \frac{1}{4} i \omega' = \frac{1}{4} i q \sinh \beta
}
(${\rm Re} s < 0$), and demand that $\Psi(u)$ is analytic at $u=1$. 
This works whenever
\eqn{nothalfint}{
2 s \neq \mbox{integer} \; .
}
Indeed, the regular solution behaves as $(1-u)^{-s}$, and the corresponding
$\Psi$ as $(1-u)^{-2s}$. Provided the inequality \eno{nothalfint} is satisfied,
this is not analytic
and is rejected by our boundary condition.

Note that the inequality \eno{nothalfint} is sufficient but not necessary 
for the singular
boundary problem to make sense. Suppose \eno{nothalfint} is
not satisfied for some values of $q$ and $\beta$, but both solutions for $\Psi$ 
are regular at $u = 1$. We consider such a $q$ to be an eigenvalue of our problem 
(at that particular $\beta$), because we can always form a linear combination of the
two regular solutions that satisfies the second boundary condition \eno{bc0}. 
On the other hand, it is a priori possible that \eno{nothalfint} 
fails in such a way that one of the solutions for $Z$ contains a logarithm of $1-u$. 
In this case, we truly have no recourse; indeed, the singular part cannot even be
peeled off as in \eno{peel}. Interestingly, the condition (\ref{nothalfint}) never 
breaks down for the ``main'' branch of $q(v)$, as defined below.

We anticipate that there is more than one eigenvalue of $q$ for each value
of the shock's speed $v$. We refer to these as 
different branches and denote them as $q_n(v)$. 
Let us mention some of the
properties of these eigenvalues for the case $\omega=0$.

First, setting $q=-i \kappa$ makes all the 
coefficients in the equation (\ref{QN2}) real and turns the condition (\ref{bc1}) 
real as well. We conclude
that the eigenvalues $q_n(v)$ are all purely imaginary, and $s$ all purely real. 
(This is not the case at $\omega\neq 0$.) 

Second, while the functions
(\ref{P}) and (\ref{Q}) do not depend on the sign of $q$, the condition (\ref{bc1})
does. Hence, the set of the eigenvalues $q_n(v)$ at a nonzero $v$ 
is not symmetric about $q=0$. The reflection $q\to -q$, without changing the direction
of the shock's velocity, is equivalent to reflection of both
space and time: $x\to -x$ and $t\to -t$. In particular, it exchanges the subsonic
and supersonic sides of the shock. The absence of symmetry under $q\to -q$
corresponds to the condition (already noted in Sec.\eno{VISCHYDRO}) that the fluid 
must be supersonic in front of the shock and subsonic behind it, and never vice 
versa---a condition that reflects, ultimately, the second law of thermodynamics. 
In calculations on the gravity 
side, the source of this irreversibility is the choice of the infalling wave in
\eq{infall}.

{\em The main branch.} The branch $q_0(v)$, for which $\exp(i q x)$ decays
away from the shock the slowest, will be referred to as the ``main'' branch and
often denoted simply as $q(v)$. 
This is the branch that crosses zero at $v= 1/\sqrt{3}$ and is the only one
seen in the small-gradient (hydrodynamic) approximation
discussed in Sec.~\ref{SHHYDRO}. 

{\em The exceptional case.}
The preceding discussion of the boundary conditions does not apply to 
the exceptional case $\omega=0$,
$\beta =\beta_{\rm cr}$, when the singularities at $u=1$ and $u=u_1$ coincide. 
This case needs to be considered separately.
At $\omega=0$ and $\beta =\beta_{\rm cr}$, Eq.~(\ref{QN2}) becomes
\eqn{QN2crit}{
Z'' + \frac{3u^2 -1}{u f(u)} Z' + \frac{1}{f^2(u)}\left(
4 u^2 + q^2 \frac{3u^2 -1}{4u} \right) Z = 0 \, .
}
Solutions near $u=1$ are of the form $Z(u) \sim (1-u)^s$ with
\eqn{scrit}{
s = 1 \pm i \frac{q}{2\sqrt{2}} \, .
}
Consider the substitution
\eqn{substPhi}{
Z(u) = (1-u^2)^{1/2} \Phi(u) \, .
}
Eq.~(\ref{QN2crit}) becomes
\eqn{eqnPhi}{
\Phi'' - \frac{1}{u} \Phi' + \frac{4 u^3 + q^2 (3 u^2 -1)}{4 u f^2(u)} \Phi 
= 0 \, .
}
Note that for $q^2 = -2$ the coefficients in \eq{QN2crit} are all regular at 
$u=1$. Hence, $q=\pm i \sqrt{2}$ are eigenvalues of the boundary problem.
Since $\beta = \beta_{\rm cr}$ is supersonic, only 
\eqn{qcrit}{
q = - i \sqrt{2} 
}
is physical; it lies on the main branch. Curiously, although
the behavior at $u=1$ prescribed by \eq{scrit} is in general different from
that prescribed by \eq{bc1}, for $q = - i \sqrt{2}$ (and $s = 1/2$) they 
coincide. As a result, the curve $q(v)$ corresponding to the main branch
is continuous at $v = v_{\rm cr}$. 

For branches above the main branch, $i q_n(v_{\rm cr})$ is large, so that only 
one of the solutions to \eq{QN2crit} is regular at $u=1$. 
The boundary condition is to choose the regular solution, which corresponds to 
choosing the plus sign in \eq{scrit}. This is equivalent to using 
\eqn{peelcr}{
Z(u) = (1-u)^{i q / 2 \sqrt{2}} \Psi(u) 
}
[cf. \eq{bc1}] but allowing $\Psi(u)$ to vanish (linearly) at $u=1$.
Indeed, as $v$ goes through $v_{\rm cr}$, $\Psi(1)$ goes continuously through
zero. 
Thus, these other branches are also continuous at $v = v_{\rm cr}$.

\subsection{Quasinormal modes for special values of $v$}

\subsubsection{$v=0$ (fluid at rest)}

The case of a plasma at rest has already been studied. In particular, as argued for example in \cite{Danielsson:1998wt},
the eigenvalue $q$ is given by the lowest glueball mass in Witten's $QCD_3$ construction \cite{Witten:1998zw}. The reason is
that, in the Euclidean space,  both finite temperature $\mathcal{N}=4$ and $QCD_3$ are dual to the same $AdS$ black hole. For the channel 
we are considering, the glueball mass was computed in \cite{Brower:2000rp} giving $iq=2.3361$ in perfect agreement with our
numerical results. This provides a nice check of the calculation although we should point out that, in the frame where the 
shock wave is at rest, the velocity of the fluid is always $v>\frac{1}{3}$ so $v=0$ is not directly relevant to our problem. 

\subsubsection{$v=1/\sqrt{3}$ (the speed of sound)}

This is the limit when the strength of the shock (as measured by changes in various
quantities between the left and right of the shock) vanishes. As we have seen
in section \eno{SHHYDRO}, in this limit $q=0$. The corresponding solution to \eq{QN2} is
\[
Z(u) = u^2 \, .
\]
\subsubsection{$v=\sqrt{2/3}$ (the singular point)}

In this case the pole corresponding to the horizon $u=1$ merges with the one at $u=u_1$ since
$v=\sqrt{2/3}$ implies $u=u_1$. Interestingly in this case we can find the exact eigenvalue $iq=\sqrt{2}$
and the eigenfunction is given in terms of a hypergeometric function:
\beq
y(u) = u \sqrt{1-u^2}   \left(\frac{1+u}{u}\right)^{\frac{1+i}{2}}  F_{2,1}\left(\frac{3+i}{2}, \frac{-1+i}{2} ; 1+i; 1+\frac{1}{u}\right)\;.
\eeq
With this definition we find
\beq
Z(u) = -\mbox{Re}(y(u)) + C_1 \mbox{Im}(y(u))\;,
\eeq
where $C_1$ is a constant that we evaluate numerically to be $C_1=0.38898$ from the boundary conditions. 

\subsubsection{$v\rightarrow 1$ (ultrarelativistic limit)}\label{ULTRA}
Numerical solution (described in the next subsection) shows that the values of $s$, 
\eq{sdef}, for the physical branches become large in the limit $v\to 1$, and the
maxima of the eigenfunctions scale towards $u=0$. This 
suggests that we can obtain an equation applicable in the
ultrarelativistic limit by neglecting $u$ in comparison with unity 
in the coefficient functions (\ref{P}) and (\ref{Q}). We obtain
\eqn{Prel}{
P(u)  =  \frac{3 - 5 \g^2 u^2}{u (\g^2 u^2 - 3)} \;,
}
\eqn{Qrel}{
Q(u)  =  - \frac{4 \g^2 u^2}{\g^2 u^2 - 3} + q^2 \frac{\g^2 u^2 - 1}{4u}\;.
}
These expressions suggest further that we define a new variable $x$, 
as follows:
\be
x = \frac{u^2}{u_1^2} = \frac{1}{3} u^2 \cosh^2 \beta \, , 
\label{scaled_x}
\ee
and take the formal limit $\cosh\beta \to \infty$ while keeping $x$ fixed.
\eq{QN2} becomes
\be
Z'' + \frac{2}{1-x} Z'
+ \frac{p^2}{16} \left( 3 - \frac{1}{x} \right) \frac{1}{\sqrt{x}}Z = 0 \, ,
\label{scaled_eq}
\ee
where primes now denote derivatives with respect to $x$, and
\be
p^2 \equiv  q^2 u_1 = \frac{q^2 \sqrt{3}}{\cosh\beta} \, . 
\label{scaled_q}
\ee
The change of variables (\ref{scaled_x}) and the limit $\cosh\beta \to \infty$ 
map the physical region $0 < u < 1$ to $0 < x < \infty$, with the Dirichlet 
boundary 
conditions at both ends. As before, the point $x = 1$ (formerly $u = u_1$) is
a singular point of the equation but not of either of the two linearly 
independent solutions. Thus, in numerical integrations we can circumvent this
point by first displacing it into the complex plane, i.e., replacing $1-x$ in 
with $1 - x + i \epsilon$ in \eq{scaled_eq} (the sign of $\epsilon$ does not 
matter), and then taking the limit of the solution at $\epsilon \to 0$. 

\begin{figure}
\begin{center}
\includegraphics[scale=1.]{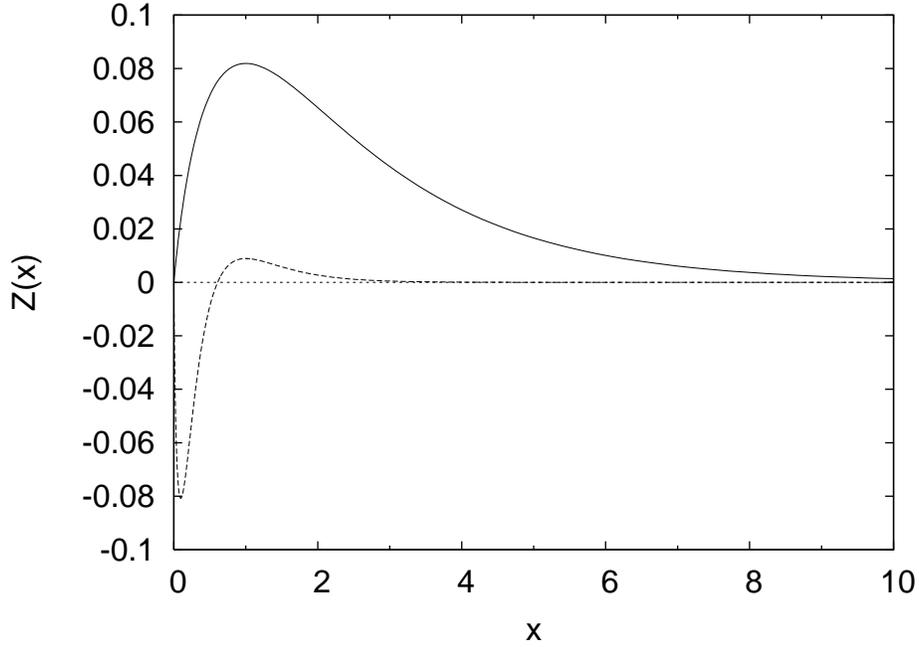}
\end{center}
\caption{The wavefunctions (arbitrarily normalized) corresponding to the ground state 
(solid line) and the first excited state (dashed line) of \eq{scaled_eq}.}
\label{fig:scaled_sol}
\end{figure}

It follows from \eq{scaled_eq} that all solutions must have 
extrema (maxima, if we agree to choose the overall sign of $Z$ in a certain way)
at $x=1$. The solutions vanish linearly at $x=0$ and, provided $\imp < 0$, 
exponentially at $x\to \infty$:
\be
Z(x\to \infty) \sim x^{9/8} \exp( - i \frac{1}{\sqrt{3}} p x^{3/4}) \, .
\label{scaled_asymp}
\ee
In Fig.~\ref{fig:scaled_sol}, we plot the eigenfunctions corresponding to the
smallest two values of $|\imp|$. According to \eq{scaled_q}, these determine the 
asymptotics of the main (lowest) and the next lowest branches of $q_n(v)$ in the
ultrarelativistic limit $v \to 1$. Numerically, we find
\ba
i q_0(v) & = & 1.895 \sqrt{\gamma} \, , \label{q0_asymp} \\
i q_1(v) & = & 5.424 \sqrt{\gamma} \, . \label{q1_asymp}
\ea
Note that numerical solution is needed only to determine the coefficients 
in these formulas:
the scaling with $\gamma$ follows directly from \eq{scaled_q} and 
the fact that \eq{scaled_eq} contains no parameters.

\subsection{Numerical results}
Apart from the very few values of $v$ (discussed earlier)
for which we have found analytical solutions to \eq{QN2}, we have resorted
to solving this equation numerically. We have used two numerical methods: 
(i) the shooting method and (ii) the series expansion. Where their domains 
of applicability overlap, these methods have produced equivalent results.

In the shooting method, we peel off the non-analytic part as in \eno{peel} and
set up an initial value problem for $\Psi(u)$ at $u$ close to 1. From 
\eno{QN2}, the expansion of $\Psi(u)$ near $u=1$ is
\eqn{Zexp1}{
Z(u\to 1) = (1-u)^s \left[ 1 + A (1-u) + \ldots \right] \, ,
}
where $s$ is given by \eno{sdef} and
\eqn{Acoef}{
A = \frac{1}{8(2s+1)} \left( {q'}^2 - {\omega'}^2 - 4 s 
+ \frac{16 {q'}^2 (2s -1)}{2 {q'}^2 - 3 {\omega'}^2} \right) \, .
}
The initial value problem is
\begin{eqnarray}
\Psi(1-\delta) & = & 1 \, , \label{Psi1} \\
\Psi'(1-\delta) & = & - A \, .
\label{Psipr1}
\end{eqnarray}
We can then adjust $q$ (on which both $q'$ and $\omega'$ depend) so that
the boundary condition \eno{bc0} at the other end is satisfied. The limit
$\delta \to 0$ is expected to be smooth whenever the boundary conditions
\eno{Psi1}--\eno{Psipr1} are sufficient to reject the second solution.
This is always the case for the regular problem (${\rm Re} s > 0$), but not
for the singular one (${\rm Re} s < 0$). In the latter case, 
\eno{Psi1}--\eno{Psipr1} are sufficient only if
\eqn{scond}{
\mbox{Re} s > - \frac{1}{2} \, ,
}
which is a stronger condition than (\ref{nothalfint}).

Another caveat is that we need
to develop a way for circumventing the singular point $u=u_1$, in the case 
when the shock velocity (relative to the fluid) exceeds the critical value
given by (\ref{betacr}), and the singularity moves into the physical 
region $0<u< 1$. Even though, as we have seen, solutions to \eq{QN2} are 
always regular  at $u=u_1$, the singularity in the coefficient functions 
precludes passing through this point by means of a numerical integration.
The approach we adopt here is to consider solutions that are not exactly
static in the boosted frame but oscillate with a small (real) frequency 
$\omega$. A nonzero $\omega$ displaces the singularity into the complex 
plane, so that the equation can be integrated numerically. The eigenvalues
and eigenfunctions at $\omega=0$ can then be obtained as limits of those
at $\omega\neq 0$ as $\omega\to 0$. The absence of singularity in the 
solutions guarantees that these limits are smooth.

In the series expansion method, one develops two power series expansions,
one near $u=0$, the other near $u=1$,
starting with the terms prescribed by the boundary conditions
(\ref{bc0}) and (\ref{bc1})
(after peeling off the non-analytic behavior at $u\to 1$ as in 
(\ref{peel})). The logarithmic derivatives of these two expansions are then 
matched at an intermediate
point of the interval $0<u< 1$. One expects that, if the expansions near
the endpoints are taken to sufficiently high orders, the results will be
insensitive to the precise value of $u$ at which the matching occurs.
This method does not require any special device to circumvent the singular
point $u=u_1$.  Indeed, that can be verified by developing a third series around the point $u=u_1$ and then matching the logarithmic derivatives with the two series developed around $u=0$, $u=1$.  The results are indistinguishable numerically.

In Fig. \eno{Qbranches} we show several branches of $q_n(v)$ obtained by these methods. It is interesting to note that a good approximation to the main 
branch is given by \footnote{Here we restore the dependence on $T$.}
\eqn{GOTFIRSTFIT}{
 \frac{iq}{\pi T} = 4\left(\frac{3}{2}\right)^{\frac{1}{4}}\sqrt{\g}\left(v-\frac{1}{\sqrt{3}}\right)\;.
}
A better approximation can be found by including more parameters in the fit.  Including one more parameter, the curve 
\eqn{GOTQHFIT}{
\frac{iq}{\pi T}= \sqrt{\g}\frac{1-v\sqrt{3}}{Fv+G} \;,
}  
where 
\eqn{GOTFITPAR}{
F= \frac{1}{2}\left(\frac{3}{2}\right)^{3/4}(2^{1/4}-1)^2 \;,\quad G=-\frac{3^{1/4}(1+2^{1/2}-2^{3/4})}{2^{5/4}} \;,
}
gives a better approximation close to the speed of sound and the large $\g$ asymptotics. 

These approximations can also be used to obtain an approximate function $\omega'(q')$ 
(the dispersion law) for the sound waves.  Transforming to the unboosted frame and using 
$v=-\frac{\oo'}{q'}$, we find that the approximation \eno{GOTFIRSTFIT} provides us with 
the following implicit equation for $\omega'(q')$:
\eqn{GOTFITDSP}{
q'(q'^2-\oo'^2)^{3/2}=16 \left(\frac{3}{2}\right)^{\frac{1}{2}}\left(\oo'+\frac{q'}{\sqrt{3}}\right)\;.}
The other fit can also be used in this way, but the resulting equation is more complicated and we omit it here.

\begin{figure}
\begin{center}
\includegraphics[width=12cm]{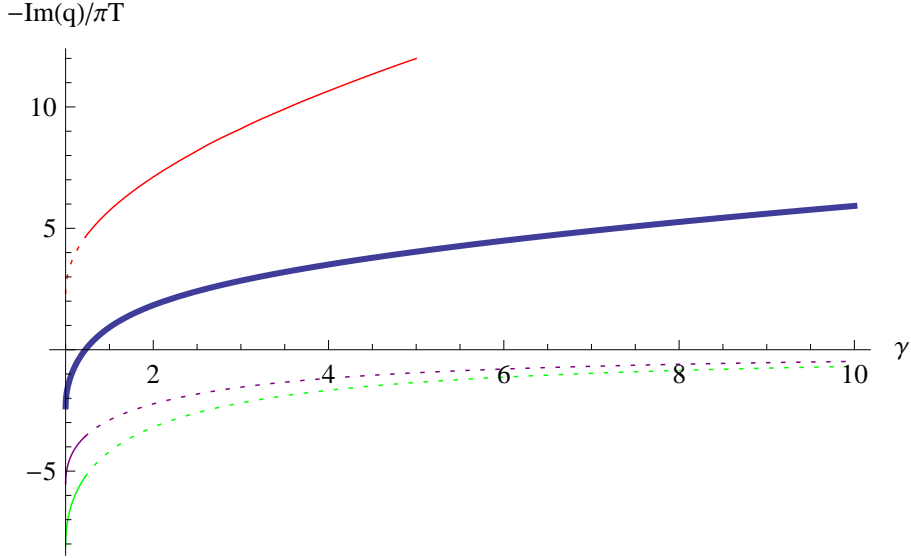}
\end{center}
\caption{Imaginary parts of $q/(\pi T)$ versus $\g=\frac{1}{\sqrt{1-v^2}}$ are plotted.  The main branch crosses $0$ at the speed of 
sound. The dotted sections of the curves denote unphysical values of $q$---those that are 
discarded as they
correspond to perturbations growing exponentially at the respective asymptotic infinities.
The solid sections denote physical values (corresponding to $-\Im (q)<0$ for 
subsonic $\gamma$ and $-\Im(q)>0$ for supersonic). The asymptotics for $\g \rightarrow \infty$ for the main branch and the one above it are discussed 
in section \eno{ULTRA}. }
\label{Qbranches}
\end{figure}

\section{Discussion}

 In this work we have used the AdS/CFT correspondence to study shock waves 
propagating in a strongly coupled plasma. Shock waves appear quite 
generically when the motion of
the fluid is supersonic and produce dissipation and drag even for zero 
viscosity. In the case of ideal fluids they are associated with surfaces 
where the velocity and pressure
are discontinuous. The discontinuity indicates a failure of the 
hydrodynamic approximation and should generically be resolved by a 
microscopic description of the system which,
in this context, is provided by the dual gravity description. An 
exception is the case of weak shocks, which propagate close 
to the speed of sound, where the inclusion of
dissipation, namely, viscosity, resolves the shock. In this case, the 
dual metric can be found using an expansion in the strength of the 
shock. On the other hand, strong shocks are
beyond the hydrodynamic approximation. They can only be resolved by 
finding the dual gravity solution which should be a smooth wave 
propagating without deformation on the horizon
of the black hole. Far from the shock, the solution differs slightly 
from a boosted black hole which allows for a perturbative study of the 
solution. In particular, we have computed,
in the rest frame of the shock, the exponential tail of the solution, 
namely, the width or penetration depth of the shock. It is a function of 
the velocity, which we have determined numerically. In particular,
when the speed of the incoming fluid approaches the speed of light, the 
penetration depth ahead of the shock goes to zero as the inverse square 
root of the gamma factor,
$\ell\sim \gamma^{-\half}$. Since the length scale goes to zero, this 
scaling exponent is an ultraviolet property of the theory, as can also 
be seen from the bulk calculation, where
the exponent is determined by the properties of the metric near the 
boundary. It would be of interest for future work to establish the value 
of this exponent for
other backgrounds in the context of AdS/CFT or perhaps even directly from 
perturbative gauge theory calculations, for example, in QCD. More 
generically, since shock waves probe
microscopic properties of the system, they are an ideal tool to study the 
transition from the microscopic to an effective hydrodynamic 
description. For example, we have shown that,
for strong shocks, the dependence of the penetration depth on the 
velocity of the incoming fluid is not correctly reproduced either by 
second order hydrodynamics or by the
Israel-Stewart theory. It is possible to encode our results into effective linearized hydrodynamics of the type proposed in \cite{Lublinsky:2009kv}, but with effective viscosity given by a numerically determined function of $\oo'$ and $q'$.  The linearized description, however, is valid 
only far from the shock. It would be interesting to see if an improved 
effective description exists
that can correctly capture the main properties of shock waves in the nonlinear
region.

 Although we have understood several basic properties of shock waves in 
the context of AdS/CFT, there are many interesting questions that we 
have not addressed here and
would be interesting to pursue. One important question is if the full 
solutions dual to shock waves can be found analytically or by 
numerical methods. They are interesting
objects in gravity since they correspond to black branes with different 
asymptotic temperatures on the two sides of the wave. Such waves propagate 
without deformation and generate
entropy by expanding the area of the horizon. Perhaps they are quite a 
generic phenomenon not restricted only to examples appearing in the 
context of AdS/CFT. 
 Other, perhaps simpler problems to consider are related to the 
introduction of dynamical quarks by means of probe branes \cite{Bigazzi:2009bk} in the 
background of the shock.  A shock should appear
on the brane giving rise to a force on quarks and meson emission from 
the shock. Finally, the introduction of dynamical quarks can also 
provide a closer point of contact with the quark-gluon 
plasma experiments at RHIC where generation of a Mach cone by 
a heavy quark propagating in the plasma has been recently suggested 
\cite{Abelev:2008nda}.   For that reason, it would be
of great interest to understand the conditions under which a moving quark 
generates a strong shock such as the one studied in the present paper.

\section{Acknowledgements}

We are grateful to Denes Molnar and Fuqiang Wang for suggestions and discussions. This work was supported in part by DOE under grant 
DE-FG02-91ER40681. The work of M.K. was also supported in part by the Alfred P. Sloan Foundation and by NSF under grant 
PHY-0805948.

\clearpage

\appendix

\section{The equations of motion for the perturbations}\label{EOMS}

In this appendix we derive the equations of  motion for various perturbations of the metric.  The background metric is given by
\eqn{BOOSTBH}{
ds^2_0 = r^2 \eta_{\mu\nu} dx^\mu dx^\nu 
+ \frac{r_0^4}{r^2} (dt \frac{1}{\sqrt{1-v^2}} - dx \frac{v}{\sqrt{1-v^2}})^2 
+ \frac{dr^2}{r^2 (1 - r_0^4/r^4)} \, ,
}
and the perturbation by 
\eqn{BOOSTBHPERT}{
ds^2_1 = r^2 e^{iq x}\left( H_{00} dt^2 + H_{11} dx^2 + 2 H_{01} dt dx + H (dy^2 + dz^2)
\right)\;.
}
  There are seven independent equations coming from
\eqn{PERTEINSTEIN}{
\tilde{G}_{MN}=R_{MN}+4g_{MN}=0\;.
}
We form the linear combination 
\eqn{LINEARCOMBEIN}{
\mathcal{L}=A^{MN}\tilde{G}_{MN}\;,
}
where the only non zero entries for the $A$ matrix are the linearly independent equations, which are the $rr,rt,rx,tt,xx,yy,xt$ components of $\tilde{G}_{MN}$ .  After choosing four entries for $A_{MN}$, namely $A_{rr},A_{rx},A_{xx},A_{yy}$, one can eliminate $H_{01},H_{11}$ and their derivatives from  $\mathcal{L}$.  Only four constants are needed since two of the equations are first order.  After this operation $\mathcal{L}$ is a function of only $H_{00},H$ and their derivatives.  One cannot use the three remaining constants to eliminate one of the functions and it's derivatives.  The reason is that only two constants are free, the third one can be thought of as an overall rescaling of $\mathcal{L}$ and there are three coefficients to eliminate, the three factors multiplying $H,H',H''$.  A redefinition
\eqn{ZDEF}{
H_{00}(r)=Z(r)-g(r)H(r)\;,
}   
and choosing $g(r)$ such that the coefficient of $H''$ vanishes allows us to write a decoupled equation for $Z(r)$.  With the choice of 
\eqn{GOTG}{
g(r)=1+\frac{r_{h}^4}{r^4}\frac{1}{1-v^2}\;,
} 
the final equation for $Z(r)$ is 
\eqn{EOMZ}{
Z''(r)+ Z'(r) &\frac{1}{r}\left( \frac{5r^4-r_{h}^4}{r^{4}-r_{h}^4} +\frac{8r_{h}^{4}}{r_{h}^4-3r^4(1-v^2)}\right)-\cr 
&- Z(r)\left( \frac{16r_{h}^8}{r^2(r^4-r_{h}^4)(r_{h}^4-3r^4(1-v^2))}-q^2 \frac{r_{h}^4-r^4(1-v^2)}{(r^4-r_{h}^4)^2(1-v^2)}\right)=0\;. 
}
This equation coincides with the equation for the sound pole \cite{Kovtun:2005ev} when one boosts to the frame where the black hole is moving.  Now we can trace back the equations and find the equations of motion for the rest of the perturbation components.  Tracing back the procedure to derive the equation for $Z(r)$ we find that 
\eqn{GOTHEOM}{
H'(r)= H(r)\frac{2r_{h}^4}{r(r^4-r_{h}^4)}-Z'(r)\frac{r^4(1-v^2)}{r_{h}^4-3r^4(1-v^2)}+Z(r) \frac{2r^3r_{h}^4(1-v^2)}{(r^4-r_{h}^4)(r_{h}^4-3r^4(1-v^2))}\;,
}
\eqn{GOTH00EOM}{
H_{00}'(r)=H_{00}(r)&\frac{2r_{h}^4(r^4(1+v^2)-3r_{h}^4)}{r^5v^2(r^4-r_{h}^4)-r(r^8-r_{h}^8)}+Z'(r)\left(\frac{2}{3}+\frac{4r_{h}^4}{3\left( r_{h}^4-3r^4(1-v^2)\right)}\right)+\cr
                   &+Z(r)\frac{4r_{h}^4\left( -2r_{h}^8+r^4r_{h}^4(4-3v^2)+r^8(v^2+v^4-2)\right)}{r(r^4-r_{h}^4)(r_{h}^4+r^4(1-v^2))(r_{h}^4-3r^4(1-v^2))}\;.
}
We can treat the $Z(r),Z'(r)$ terms as a source since $Z(r)$ satisfies a decoupled equation.  For the last two perturbations $H_{01},H_{11}$ it is easier to define a linear combination of them
\eqn{TILDEHDEF}{
\tilde{H}(r)=vH_{11}(r)+2H_{01}(r)\;.
}
Treating terms containing $H(r),Z(r)$ as sources $\tilde{H}$  satisfies
\eqn{TILDEHEOM}{
\tilde{H}'(r)=&\tilde{H}(r)\frac{4r_{h}^4}{r(r^4-r_{h}^4)}+Z'(r)\frac{2r_{h}^4(r^4-r_{h}^4)v\left(r_{h}^4(v^2-2)+4r^4(1-v^2)\right)}{(r^4-r_{h}^4)(r_{h}^4-3r^4(1-v^2))^2}-\cr
                        &-Z(r)\frac{2vr_{h}^4\left(3r_{h}^8(v^2-2)-9r^8(1-v^2)^2+r^4r_{h}^4(17-20v^2+3v^4)\right) }{r(r^4-r_{h}^4)(r_{h}^4-3r^4(1-v^2))^2} +\cr 
                         &+Z(r)\frac{vq^2r^6(1-v^2)}{(r^4-r_{h}^4)(r_{h}^4-3r^4(1-v^2))} -\cr
                         &-H(r)\frac{v\left(q^2r^6(v^2-1)+2r_{h}^4\left(r^4(5-3v^2)+r_{h}^4(v^2-2)\right)\right)}{r^5(r^4-r_{h}^4)(1-v^2)}\;.}
Having determined $\tilde{H}(r)$ the last two perturbations satisfy
\eqn{H01EOM}{
H_{01}'(r)=&\tilde{H}(r)\frac{2r_{h}^4}{r(r^4-r_{h}^4)(1-v^2)}+H(r)\frac{2vr_{h}^4\left(r_{h}^4(1-2v^2)-3r^4(1-v^2) \right)}{r^5(r^4-r_{h}^4)(1-v^2)^2}-\cr
                     &-Z'(r)\frac{2vr_{h}^4}{r_{h}^4-3r^4(1-v^2)}-Z(r)\frac{2vr_{h}^4\left( 3r^4(1-v^2)+r_{h}^4(2v^2-3)\right)}{r(r^4-r_{h}^4)(1-v^2)(r_{h}^4-3r^4(1-v^2))} \;,
}
\eqn{H11EOM}{
H_{11}'(r)=&-\tilde{H}(r)\frac{4vr_{h}^4}{r(r^4-r_{h}^4)(1-v^2)}+Z'(r)\frac{2r_{h}^4\left( v^2r_{h}^4-2r^4(1-v^2)\right)}{(rh_{h}^4-3r^4(1-v^2))^2}+ \cr 
                 &+H(r)\left(\frac{q^2r}{r^4-r_{h}^4}+\frac{2r_{h}^4\left(v^2r_{h}^4(1+v^2)+r^4(1+2v^2-3v^4) \right)}{r^5(r^4-r_{h}^4)(1-v^2)^2}\right) +\cr 
                  &\;+Z(r)\frac{2r_{h}^4\left( -9r^8(1-v^2)^2(1+v^2)+v^2r_{h}^8(3v^2-5)+r^4r_{h}^4 (7+v^2-11v^4+3v^6)\right)}{r(r^4-r_{h}^4)(1-v^2)(r_{h}^4-3r^4(1-v^2))^2} + \cr
                   &\;\;+Z(r)\frac{q^2r^5(1-v^2)}{(r^4-r_{h}^4)(r_{h}^4-3r^4(1-v^2))} \;.
}  
We can now find the asymptotic behavior for all perturbations close to the horizon and close to the boundary.  The results are summarized in table \eno{TABLEWAVE}.
\begin{table}[ht]\label{TABLEWAVE}
\begin{center}
\begin{tabular}{||c || c | c | c|}
\hline
   & $r\rightarrow \infty $&$r\rightarrow r_h$& $r\rightarrow r_{f}$  \\
\hline
 $Z(r)$  & $r^{-4}$ & $(r-r_{h})^{-i \frac{qv}{4r_{h} \sqrt{(1-v^2)}}} $ & $\z_{0} + \z_{1}(r-r_{f}) +\z_{2}(r-r_{f})^2+\cdots$ \\
\hline
  $H_{00}(r)$ & $r^{-4}$ &$\frac{2v^2}{3v^2-2} (r-r_{h})^{-i \frac{qv}{4r_{h} \sqrt{(1-v^2)}}}$, $v< \sqrt{\frac{2}{3}}$  &$h^{(0)}_{0}+h^{(0)}_{1}(r-r_{f})+\cdots$\\
\hline
  $H(r)$ & $r^{-4}$ & $\frac{1-v^2}{3v^2-2}(r-r_{h})^{-i \frac{qv}{4r_{h} \sqrt{(1-v^2)}}} $, $v<\sqrt{\frac{2}{3}}$ &$h_{0}+h_{1}(r-r_{f})+\cdots$\\
               &                &    $\tilde{C}_{1}  \sqrt{r-r_h} $, $v\ge \sqrt{\frac{2}{3}} $                                                                         &  \\
\hline
  $\tilde{H}(r)$ & $r^{-9}$ &  $-\frac{2iv(-iqv+2(1-2v^2)\sqrt{1-v^2}-q^2(1-v^2)^{3/2})}{(3v^2-2)(qv-4i\sqrt{1-v^2})}\cdot$&$\tilde{h}_{0} +\tilde{h}_{1}(r-r_f)+\cdots$ \\
           & & $\cdot(r-r_{h})^{-i \frac{qv}{4r_{h} \sqrt{(1-v^2)}}}$,  $v<\sqrt{\frac{2}{3}}$  & \\
           & &  $\tilde{C}_{2} \sqrt{r-r_{h}}$ ,$v\ge \sqrt{\frac{2}{3}}$                   &\\
\hline
  $H_{01}(r)$ & $r^{-9}$ &  &$h^{(01)}_{0}+h^{(01)}_{1}(r-r_{f})+\cdots $ \\
\hline
  $H_{11}(r)$ & $r^{-9}$ &  &$h^{(11)}_{0}+h^{(11)}_{1}(r-r_{f})+\cdots $\\
\hline
\end{tabular}
\caption{The behavior of the perturbations close to the boundary, the horizon and the pole of the equation \eno{EOMZ} is given.  Here, $r_{f}=\frac{1}{3^{1/4}}\frac{r_{h}}{(1-v^2)^{1/4}}$ is the location of the pole of \eno{EOMZ}.  The boundary conditions for the perturbations at the boundary $r\rightarrow \infty$ are that the metric is unchanged from the Minkowski metric. At the horizon the condition is that the asymptotic behavior corresponds to an infalling graviton in the unboosted black hole. In the linearized approximation the overall scaling factor does not appear and we only show the power law behavior. The normalization of the perturbations close to the horizon are relative to the normalization of $Z(r)$.  The normalization of $Z(r)$ is taken to be $1$ for the factor multiplying $(r-r_{h})^{-i \frac{qv}{4r_{h} \sqrt{(1-v^2)}}} $. For $v\ge \sqrt{\frac{2}{3}}$ the relative coefficients $\tilde{C}_{1,2}$ are not computed.  The asymptotic behavior of $H_{01},H_{11}$ is not shown for brevity but can be easily inferred from \eno{H01EOM}-\eno{H11EOM}.    }
\end{center}
\end{table}

The behavior of $H_{01}$ and $H_{11}$ close to the boundary are consistent with the equations of motion for the boundary stress energy tensor.  The last can be rewritten as 
\eqn{STRESSEOMB}{
 \d T^{01}=0\;,\quad  \d T^{11}=0\;,
}
where $\d T^{\m\n}$ denote the perturbations away from the ideal boosted fluid stress energy tensor.  From the $AdS/CFT$ dictionary we know that 
\eqn{GOTBOUNDSTR}{
\d T^{01}=\lim_{r\rightarrow \infty} r^2 H^{01}(r)=0\;, \quad \d T^{00}= \lim_{r \rightarrow \infty} r^2 H^{11}(r)=0\;.  
} 
in agreement with \eno{STRESSEOMB}.

\section{Expansion near the boundary}

Given a conserved boundary energy momentum tensor the boundary conditions for an asymptotic AdS metric are fixed. There is a procedure \cite{deHaro:2000xn} that allows one to find such metric expanded in powers
of $\frac{1}{r}$ where $r$ is the radial coordinate in the Poincare AdS patch, such that $r=0$ is the horizon and $r=\infty$ the boundary. In our case the procedure simplifies. We fix the energy 
momentum tensor to be  
\beq
T^{tt}= \ve(x), \ \ T^{tx}=C_1, \ \ T^{xx}=C_2, \ \ T^{yy}=T^{zz}=\half \ve-\half C_2\;,
\eeq
which is obviously conserved ($\partial_\mu T^{\mu\nu}=0$). The energy density $\ve(x)$ has to be computed from the hydrodynamic equations or a guess can be made. In any case this fixes the boundary 
condition and allows us to extend the metric as:
\beqa
g_{tt} &=& \framebox{\(\displaystyle -r^2 - \frac{\ve}{r^2}\)}+\frac{\ve''}{12r^4} + \frac{1}{r^6}\left(\frac{\ve C_2}{24}-\frac{7\ve^2}{16}+\frac{C_2^2}{16}+\frac{5C_1^2}{48}-\frac{\ve^{iv}}{384}\right) \nonumber \\
      &&   +\frac{1}{r^8}\left(\frac{3}{40}\ve\ve''-\frac{7}{360}C_2\ve''-\frac{1}{480}\ve'{}^2+\frac{1}{23040}\ve^{vi}\right)
          +\frac{1}{r^{10}} \left(\frac{C_2^2\ve}{48}-\frac{29\ve''{}^2}{9216}+\frac{77C_2\ve^{iv}}{69210}-\frac{11\ve\ve^{iv}}{4608} \right.\nonumber \\
      &&   \left.-\frac{\ve^{viii}}{2211840} +\frac{11C_1^2\ve}{144}+\frac{C_1^2C_2}{24}+\frac{C_2\ve^2}{72}-\frac{7\ve^3}{48}+\frac{\ve'\ve'''}{4608}\right)+{\cal O}\left(\frac{1}{r^{12}}\right)\;,
\eeqa
\beqa
2g_{tz} &=&\framebox{\(\displaystyle \frac{C_1}{r^2} \)} + \frac{C_1(C_2+\ve)}{2r^6}-\frac{1}{20}\frac{C_1\ve''}{r^8}+\frac{1}{r^{10}}\left(\frac{C_1\ve^{iv}}{576}-\frac{5C_1^3}{144}
           +\frac{7C_1C_2^2}{48}+\frac{11C_1C_2\ve}{72}+\frac{7C_1\ve^2}{48}\right) \nonumber\\
       && -\frac{C_1}{161280r^{12}}\left(5\ve^{vi}+48\ve'{}^2+4512\ve\ve''+736C_2\ve''\right)  +{\cal O}\left(\frac{1}{r^{14}}\right) \;.
\eeqa
\beqa
g_{xx}&=& \framebox{\(\displaystyle r^2+\frac{C_2}{r^2}\)} + \frac{1}{r^6}\left(-\frac{C_2\ve}{24}-\frac{\ve^2}{16}+\frac{7}{16}C_2^2-\frac{5}{48}C_1^2\right) 
         +\frac{1}{r^8}\left(\frac{1}{240}\ve''C_2+\frac{1}{80}\ve\ve''-\frac{1}{160}\ve'{}^2\right) \nonumber\\
      && +\frac{1}{r^{10}}\left(-\frac{C_2^2\ve}{72}-\frac{3\ve''{}^2}{5120}-\frac{C_2\ve^{iv}}{7680}-\frac{\ve\ve^{iv}}{2560}-\frac{C_1^2\ve}{24}
         -\frac{11C_1^2C_2}{144}+\frac{7C_2^3}{48}-\frac{C_2\ve^2}{48}+\frac{\ve'\ve'''}{1536}\right) \nonumber \\
      && +{\cal O}\left(\frac{1}{r^{12}}\right) \;,
\eeqa
\beqa
g_{yy}&=&g_{zz}=\framebox{\(\displaystyle r^2-\frac{C_2}{2r^2}-\frac{\ve}{2r^2} \)}+\frac{\ve''}{24r^4}+\frac{1}{r^6}\left(\frac{5}{24}C_2\ve+\frac{\ve^2}{16}+\frac{C_2^2}{16}+\frac{C_1^2}{48}-\frac{\ve^{iv}}{768}\right) \nonumber\\
     &&          +\frac{1}{r^8}\left(\frac{1}{480}\ve'{}^2+\frac{1}{46080}\ve^{vi}-\frac{19}{720}C_2\ve''-\frac{1}{80}\ve\ve''\right)+\frac{1}{r^{10}}\left(-\frac{13C_2^2\ve}{288}
                 +\frac{5\ve''{}^2}{9216}+\frac{19C_2\ve^{iv}}{17280}\right. \nonumber \\
     && \left. +\frac{\ve\ve^{iv}}{2304}-\frac{\ve^{viii}}{4423680}-\frac{C_1^2\ve}{288}-\frac{C_1^2C_2}{288}-\frac{C_2^3}{96}-\frac{13C_2\ve^2}{288}-\frac{\ve^3}{96}-\frac{\ve'\ve'''}{4608}\right)
        +{\cal O}\left(\frac{1}{r^{12}}\right)\;,
\eeqa
where the boxed terms are fixed by the boundary conditions and the rest can be computed from solving the Einstein equations. In the absence of an exact metric for the shock wave, this expansion 
provides more information about it and could possible be used in the future as a check of given solutions and as a starting point for a numerical method. Although we show a few terms, it should be noted
that using a computer algebra program we found easily the expansion up to order $\frac{1}{r^{30}}$ although it is too lengthy to display here. These are enough terms to attempt a reconstruction of 
the metric using Pad\'{e} approximants. The condition that determines the function $\ve(x)$ then comes from demanding that the metric does not develop a singularity.  In fact, in Fefferman-Graham coordinates the $AdS$-Schwarzschild black hole metric becomes degenerate at the horizon and one cannot go beyond the horizon in these coordinates.

\clearpage
\bibliographystyle{xbib}
\bibliography{adsshock}

\providecommand{\href}[2]{#2}\begingroup\raggedright\begin{thebibliography}{10}

\bibitem{Maldacena:1997re}
J.~M. Maldacena, ``{The large N limit of superconformal field theories and
  supergravity},'' {\em Adv. Theor. Math. Phys.} {\bf 2} (1998) 231--252,
\href{http://arXiv.org/abs/hep-th/9711200}{{\tt hep-th/9711200}}.
%%CITATION = HEP-TH/9711200;%%.

\bibitem{Gubser:1998bc}
S.~S. Gubser, I.~R. Klebanov, and A.~M. Polyakov, ``{Gauge theory correlators
  from non-critical string theory},'' {\em Phys. Lett.} {\bf B428} (1998)
  105--114,
\href{http://arXiv.org/abs/hep-th/9802109}{{\tt hep-th/9802109}}.
%%CITATION = HEP-TH/9802109;%%.

\bibitem{Witten:1998qj}
E.~Witten, ``{Anti-de Sitter space and holography},'' {\em Adv. Theor. Math.
  Phys.} {\bf 2} (1998) 253--291,
\href{http://arXiv.org/abs/hep-th/9802150}{{\tt hep-th/9802150}}.
%%CITATION = HEP-TH/9802150;%%.

\bibitem{Aharony:1999ti}
O.~Aharony, S.~S. Gubser, J.~M. Maldacena, H.~Ooguri, and Y.~Oz, ``{Large N
  field theories, string theory and gravity},'' {\em Phys. Rept.} {\bf 323}
  (2000) 183--386,
\href{http://arXiv.org/abs/hep-th/9905111}{{\tt hep-th/9905111}}.
%%CITATION = HEP-TH/9905111;%%.

\bibitem{'tHooft:1973jz}
G.~'t~Hooft, ``{A planar diagram theory for strong interactions},'' {\em Nucl.
  Phys.} {\bf B72} (1974)
461.
%%CITATION = NUPHA,B72,461;%%.

\bibitem{Son:2007vk}
D.~T. Son and A.~O. Starinets, ``{Viscosity, Black Holes, and Quantum Field
  Theory},'' {\em Ann. Rev. Nucl. Part. Sci.} {\bf 57} (2007) 95--118,
\href{http://arXiv.org/abs/0704.0240}{{\tt 0704.0240}}.
%%CITATION = 0704.0240;%%.

\bibitem{Shuryak:2008eq}
E.~Shuryak, ``{Physics of Strongly coupled Quark-Gluon Plasma},'' {\em Prog.
  Part. Nucl. Phys.} {\bf 62} (2009) 48--101,
\href{http://arXiv.org/abs/0807.3033}{{\tt 0807.3033}}.
%%CITATION = 0807.3033;%%.

\bibitem{Schafer:2009dj}
T.~Schafer and D.~Teaney, ``{Nearly Perfect Fluidity: From Cold Atomic Gases to
  Hot Quark Gluon Plasmas},'' {\em Rept. Prog. Phys.} {\bf 72} (2009) 126001,
\href{http://arXiv.org/abs/0904.3107}{{\tt 0904.3107}}.
%%CITATION = 0904.3107;%%.

\bibitem{Gubser:2009md}
S.~S. Gubser and A.~Karch, ``{From gauge-string duality to strong interactions:
  a Pedestrian's Guide},'' {\em Ann. Rev. Nucl. Part. Sci.} {\bf 59} (2009)
  145--168,
\href{http://arXiv.org/abs/0901.0935}{{\tt 0901.0935}}.
%%CITATION = 0901.0935;%%.

\bibitem{Gubser:2009sn}
S.~S. Gubser, S.~S. Pufu, F.~D. Rocha, and A.~Yarom, ``{Energy loss in a
  strongly coupled thermal medium and the gauge-string duality},''
\href{http://arXiv.org/abs/0902.4041}{{\tt 0902.4041}}.
%%CITATION = 0902.4041;%%.

\bibitem{Landau:1953gs}
L.~D. Landau, ``{On the multiparticle production in high-energy collisions},''
  {\em Izv. Akad. Nauk SSSR Ser. Fiz.} {\bf 17} (1953)
51--64.
%%CITATION = IANFA,17,51;%%.

\bibitem{Gubser:2008pc}
S.~S. Gubser, S.~S. Pufu, and A.~Yarom, ``{Entropy production in collisions of
  gravitational shock waves and of heavy ions},'' {\em Phys. Rev.} {\bf D78}
  (2008) 066014,
\href{http://arXiv.org/abs/0805.1551}{{\tt 0805.1551}}.
%%CITATION = 0805.1551;%%.

\bibitem{Avsar:2009xf}
E.~Avsar, E.~Iancu, L.~McLerran, and D.~N. Triantafyllopoulos, ``{Shockwaves
  and deep inelastic scattering within the gauge/gravity duality},'' {\em JHEP}
  {\bf 11} (2009) 105,
\href{http://arXiv.org/abs/0907.4604}{{\tt 0907.4604}}.
%%CITATION = 0907.4604;%%.

\bibitem{Beuf:2009mk}
G.~Beuf, ``{Gravity dual of N=4 SYM theory with fast moving sources},'' {\em
  Phys. Lett.} {\bf B686} (2010) 55--58,
\href{http://arXiv.org/abs/0903.1047}{{\tt 0903.1047}}.
%%CITATION = 0903.1047;%%.

\bibitem{Albacete:2008vs}
J.~L. Albacete, Y.~V. Kovchegov, and A.~Taliotis, ``{Modeling Heavy Ion
  Collisions in AdS/CFT},'' {\em JHEP} {\bf 07} (2008) 100,
\href{http://arXiv.org/abs/0805.2927}{{\tt 0805.2927}}.
%%CITATION = 0805.2927;%%.

\bibitem{Aichelburg:1970dh}
P.~C. Aichelburg and R.~U. Sexl, ``{On the Gravitational field of a massless
  particle},'' {\em Gen. Rel. Grav.} {\bf 2} (1971)
303--312.
%%CITATION = GRGVA,2,303;%%.

\bibitem{Dray:1984ha}
T.~Dray and G.~'t~Hooft, ``{The Gravitational Shock Wave of a Massless
  Particle},'' {\em Nucl. Phys.} {\bf B253} (1985)
173.
%%CITATION = NUPHA,B253,173;%%.

\bibitem{Hotta:1992qy}
M.~Hotta and M.~Tanaka, ``{Shock wave geometry with nonvanishing cosmological
  constant},'' {\em Class. Quant. Grav.} {\bf 10} (1993)
307--314.
%%CITATION = CQGRD,10,307;%%.

\bibitem{Sfetsos:1994xa}
K.~Sfetsos, ``{On gravitational shock waves in curved space-times},'' {\em
  Nucl. Phys.} {\bf B436} (1995) 721--746,
\href{http://arXiv.org/abs/hep-th/9408169}{{\tt hep-th/9408169}}.
%%CITATION = HEP-TH/9408169;%%.

\bibitem{Podolsky:1997ni}
J.~Podolsky and J.~B. Griffiths, ``{Impulsive waves in de Sitter and anti-de
  Sitter space- times generated by null particles with an arbitrary multipole
  structure},'' {\em Class. Quant. Grav.} {\bf 15} (1998) 453--463,
\href{http://arXiv.org/abs/gr-qc/9710049}{{\tt gr-qc/9710049}}.
%%CITATION = GR-QC/9710049;%%.

\bibitem{Horowitz:1999gf}
G.~T. Horowitz and N.~Itzhaki, ``{Black holes, shock waves, and causality in
  the AdS/CFT correspondence},'' {\em JHEP} {\bf 02} (1999) 010,
\href{http://arXiv.org/abs/hep-th/9901012}{{\tt hep-th/9901012}}.
%%CITATION = HEP-TH/9901012;%%.

\bibitem{Emparan:2001ce}
R.~Emparan, ``{Exact gravitational shockwaves and Planckian scattering on
  branes},'' {\em Phys. Rev.} {\bf D64} (2001) 024025,
\href{http://arXiv.org/abs/hep-th/0104009}{{\tt hep-th/0104009}}.
%%CITATION = HEP-TH/0104009;%%.

\bibitem{Arcioni:2001my}
G.~Arcioni, S.~de~Haro, and M.~O'Loughlin, ``{Boundary description of Planckian
  scattering in curved spacetimes},'' {\em JHEP} {\bf 07} (2001) 035,
\href{http://arXiv.org/abs/hep-th/0104039}{{\tt hep-th/0104039}}.
%%CITATION = HEP-TH/0104039;%%.

\bibitem{Horowitz:2009fw}
W.~A. Horowitz, ``{Shock Treatment: Heavy Quark Energy Loss in a Novel AdS/CFT
  Geometry},'' {\em Nucl. Phys.} {\bf A830} (2009) 773c--776c,
\href{http://arXiv.org/abs/0907.4845}{{\tt 0907.4845}}.
%%CITATION = 0907.4845;%%.

\bibitem{Horowitz:2009pw}
W.~A. Horowitz and Y.~V. Kovchegov, ``{Shock Treatment: Heavy Quark Drag in a
  Novel AdS Geometry},'' {\em Phys. Lett.} {\bf B680} (2009) 56--61,
\href{http://arXiv.org/abs/0904.2536}{{\tt 0904.2536}}.
%%CITATION = 0904.2536;%%.

\bibitem{Friess:2006fk}
J.~J. Friess, S.~S. Gubser, G.~Michalogiorgakis, and S.~S. Pufu, ``{The stress
  tensor of a quark moving through N = 4 thermal plasma},'' {\em Phys. Rev.}
  {\bf D75} (2007) 106003,
\href{http://arXiv.org/abs/hep-th/0607022}{{\tt hep-th/0607022}}.
%%CITATION = HEP-TH/0607022;%%.

\bibitem{Gubser:2007xz}
S.~S. Gubser, S.~S. Pufu, and A.~Yarom, ``{Energy disturbances due to a moving
  quark from gauge- string duality},'' {\em JHEP} {\bf 09} (2007) 108,
\href{http://arXiv.org/abs/0706.0213}{{\tt 0706.0213}}.
%%CITATION = 0706.0213;%%.

\bibitem{Gubser:2007ga}
S.~S. Gubser, S.~S. Pufu, and A.~Yarom, ``{Sonic booms and diffusion wakes
  generated by a heavy quark in thermal AdS/CFT},'' {\em Phys. Rev. Lett.} {\bf
  100} (2008) 012301,
\href{http://arXiv.org/abs/0706.4307}{{\tt 0706.4307}}.
%%CITATION = 0706.4307;%%.

\bibitem{Chesler:2007an}
P.~M. Chesler and L.~G. Yaffe, ``{The wake of a quark moving through a
  strongly-coupled $\mathcal N=4$ supersymmetric Yang-Mills plasma},'' {\em
  Phys. Rev. Lett.} {\bf 99} (2007) 152001,
\href{http://arXiv.org/abs/0706.0368}{{\tt 0706.0368}}.
%%CITATION = 0706.0368;%%.

\bibitem{Chesler:2007sv}
P.~M. Chesler and L.~G. Yaffe, ``{The stress-energy tensor of a quark moving
  through a strongly-coupled N=4 supersymmetric Yang-Mills plasma: comparing
  hydrodynamics and AdS/CFT},'' {\em Phys. Rev.} {\bf D78} (2008) 045013,
\href{http://arXiv.org/abs/0712.0050}{{\tt 0712.0050}}.
%%CITATION = 0712.0050;%%.

\bibitem{Scheid:1974zz}
W.~Scheid, H.~Muller, and W.~Greiner, ``{Nuclear Shock Waves in Heavy-Ion
  Collisions},'' {\em Phys. Rev. Lett.} {\bf 32} (1974)
741--745.
%%CITATION = PRLTA,32,741;%%.

\bibitem{Baumgardt:1975qv}
H.~G. Baumgardt {\em et al.}, ``{Shock Waves and MACH Cones in Fast
  Nucleus-Nucleus Collisions},'' {\em Z. Phys.} {\bf A273} (1975)
359--371.
%%CITATION = ZEPYA,A273,359;%%.

\bibitem{Gutbrod:1989wd}
H.~H. Gutbrod, A.~M. Poskanzer, and H.~G. Ritter, ``{Plastic ball
  experiments},'' {\em Rept. Prog. Phys.} {\bf 52} (1989)
1267.
%%CITATION = RPPHA,52,1267;%%.

\bibitem{Gutbrod:1989gh}
H.~H. Gutbrod {\em et al.}, ``{Squeezeout of nuclear matter as a function of
  projectile energy and mass},'' {\em Phys. Rev.} {\bf C42} (1990)
640--651.
%%CITATION = PHRVA,C42,640;%%.

\bibitem{Adams:2003kv}
{\bf STAR} Collaboration, J.~Adams {\em et al.}, ``{Transverse momentum and
  collision energy dependence of high p(T) hadron suppression in Au + Au
  collisions at ultrarelativistic energies},'' {\em Phys. Rev. Lett.} {\bf 91}
  (2003) 172302,
\href{http://arXiv.org/abs/nucl-ex/0305015}{{\tt nucl-ex/0305015}}.
%%CITATION = NUCL-EX/0305015;%%.

\bibitem{Adare:2008qa}
{\bf PHENIX} Collaboration, A.~Adare {\em et al.}, ``{Suppression pattern of
  neutral pions at high transverse momentum in Au+Au collisions at $
  \sqrt{s_{NN}} = 200 GeV$ and constraints on medium transport coefficients},''
  {\em Phys. Rev. Lett.} {\bf 101} (2008) 232301,
\href{http://arXiv.org/abs/0801.4020}{{\tt 0801.4020}}.
%%CITATION = 0801.4020;%%.

\bibitem{Wang:2004kfa}
{\bf STAR} Collaboration, F.~Wang, ``{Measurement of jet modification at
  RHIC},'' {\em J. Phys.} {\bf G30} (2004) S1299--S1304,
\href{http://arXiv.org/abs/nucl-ex/0404010}{{\tt nucl-ex/0404010}}.
%%CITATION = NUCL-EX/0404010;%%.

\bibitem{Adams:2005ph}
{\bf STAR} Collaboration, J.~Adams {\em et al.}, ``{Distributions of charged
  hadrons associated with high transverse momentum particles in p p and Au + Au
  collisions at s(NN)**(1/2) = 200-GeV},'' {\em Phys. Rev. Lett.} {\bf 95}
  (2005) 152301,
\href{http://arXiv.org/abs/nucl-ex/0501016}{{\tt nucl-ex/0501016}}.
%%CITATION = NUCL-EX/0501016;%%.

\bibitem{Adler:2005ee}
{\bf PHENIX} Collaboration, S.~S. Adler {\em et al.}, ``{Modifications to
  di-jet hadron pair correlations in Au + Au collisions at s(NN)**(1/2) =
  200-GeV},'' {\em Phys. Rev. Lett.} {\bf 97} (2006) 052301,
\href{http://arXiv.org/abs/nucl-ex/0507004}{{\tt nucl-ex/0507004}}.
%%CITATION = NUCL-EX/0507004;%%.

\bibitem{Ulery:2005cc}
{\bf STAR} Collaboration, J.~G. Ulery, ``{Two- and three-particle jet
  correlations from STAR},'' {\em Nucl. Phys.} {\bf A774} (2006) 581--584,
\href{http://arXiv.org/abs/nucl-ex/0510055}{{\tt nucl-ex/0510055}}.
%%CITATION = NUCL-EX/0510055;%%.

\bibitem{Ajitanand:2006is}
{\bf PHENIX} Collaboration, N.~N. Ajitanand, ``{Extraction of jet topology
  using three particle correlations},'' {\em Nucl. Phys.} {\bf A783} (2007)
  519--522,
\href{http://arXiv.org/abs/nucl-ex/0609038}{{\tt nucl-ex/0609038}}.
%%CITATION = NUCL-EX/0609038;%%.

\bibitem{Adare:2008cqb}
{\bf PHENIX} Collaboration, A.~Adare {\em et al.}, ``{Dihadron azimuthal
  correlations in Au+Au collisions at $\sqrt{s_{NN}} = 200 GeV$},'' {\em Phys.
  Rev.} {\bf C78} (2008) 014901,
\href{http://arXiv.org/abs/0801.4545}{{\tt 0801.4545}}.
%%CITATION = 0801.4545;%%.

\bibitem{Stoecker:2004qu}
H.~Stoecker, ``{Collective Flow signals the Quark Gluon Plasma},'' {\em Nucl.
  Phys.} {\bf A750} (2005) 121--147,
\href{http://arXiv.org/abs/nucl-th/0406018}{{\tt nucl-th/0406018}}.
%%CITATION = NUCL-TH/0406018;%%.

\bibitem{Ruppert:2005uz}
J.~Ruppert and B.~Muller, ``{Waking the colored plasma},'' {\em Phys. Lett.}
  {\bf B618} (2005) 123--130,
\href{http://arXiv.org/abs/hep-ph/0503158}{{\tt hep-ph/0503158}}.
%%CITATION = HEP-PH/0503158;%%.

\bibitem{Koch:2005sx}
V.~Koch, A.~Majumder, and X.-N. Wang, ``{Cherenkov Radiation from Jets in
  Heavy-ion Collisions},'' {\em Phys. Rev. Lett.} {\bf 96} (2006) 172302,
\href{http://arXiv.org/abs/nucl-th/0507063}{{\tt nucl-th/0507063}}.
%%CITATION = NUCL-TH/0507063;%%.

\bibitem{CasalderreySolana:2004qm}
J.~Casalderrey-Solana, E.~V. Shuryak, and D.~Teaney, ``{Conical flow induced by
  quenched QCD jets},'' {\em J. Phys. Conf. Ser.} {\bf 27} (2005) 22--31,
\href{http://arXiv.org/abs/hep-ph/0411315}{{\tt hep-ph/0411315}}.
%%CITATION = HEP-PH/0411315;%%.

\bibitem{Bouras:2009nn}
I.~Bouras {\em et al.}, ``{Relativistic shock waves in viscous gluon matter},''
  {\em Phys. Rev. Lett.} {\bf 103} (2009) 032301,
\href{http://arXiv.org/abs/0902.1927}{{\tt 0902.1927}}.
%%CITATION = 0902.1927;%%.

\bibitem{Bouras:2010nt}
I.~Bouras {\em et al.}, ``{Relativistic Shock Waves and Mach Cones in Viscous
  Gluon Matter},''
\href{http://arXiv.org/abs/1004.4615}{{\tt 1004.4615}}.
%%CITATION = 1004.4615;%%.

\bibitem{LL}
L.~D. Landau and E.~M. Lifshitz, {\em Fluid Mechanics}.
\newblock Elsevier, Oxford GB, 2nd~ed., 2009.

\bibitem{Bhattacharyya:2008jc}
S.~Bhattacharyya, V.~E. Hubeny, S.~Minwalla, and M.~Rangamani, ``{Nonlinear
  Fluid Dynamics from Gravity},'' {\em JHEP} {\bf 02} (2008) 045,
\href{http://arXiv.org/abs/0712.2456}{{\tt 0712.2456}}.
%%CITATION = 0712.2456;%%.

\bibitem{Policastro:2001yc}
G.~Policastro, D.~T. Son, and A.~O. Starinets, ``{The shear viscosity of
  strongly coupled N = 4 supersymmetric Yang-Mills plasma},'' {\em Phys. Rev.
  Lett.} {\bf 87} (2001) 081601,
\href{http://arXiv.org/abs/hep-th/0104066}{{\tt hep-th/0104066}}.
%%CITATION = HEP-TH/0104066;%%.

\bibitem{Baier:2007ix}
R.~Baier, P.~Romatschke, D.~T. Son, A.~O. Starinets, and M.~A. Stephanov,
  ``{Relativistic viscous hydrodynamics, conformal invariance, and
  holography},'' {\em JHEP} {\bf 04} (2008) 100,
\href{http://arXiv.org/abs/0712.2451}{{\tt 0712.2451}}.
%%CITATION = 0712.2451;%%.

\bibitem{Natsuume:2007ty}
M.~Natsuume and T.~Okamura, ``{Causal hydrodynamics of gauge theory plasmas
  from AdS/CFT duality},'' {\em Phys. Rev.} {\bf D77} (2008) 066014,
\href{http://arXiv.org/abs/0712.2916}{{\tt 0712.2916}}.
%%CITATION = 0712.2916;%%.

\bibitem{Loganayagam:2008is}
R.~Loganayagam, ``{Entropy Current in Conformal Hydrodynamics},'' {\em JHEP}
  {\bf 05} (2008) 087,
\href{http://arXiv.org/abs/0801.3701}{{\tt 0801.3701}}.
%%CITATION = 0801.3701;%%.

\bibitem{Bhattacharyya:2008xc}
S.~Bhattacharyya {\em et al.}, ``{Local Fluid Dynamical Entropy from
  Gravity},'' {\em JHEP} {\bf 06} (2008) 055,
\href{http://arXiv.org/abs/0803.2526}{{\tt 0803.2526}}.
%%CITATION = 0803.2526;%%.

\bibitem{Israel:1976tn}
W.~Israel, ``{Nonstationary irreversible thermodynamics: A Causal relativistic
  theory},'' {\em Ann. Phys.} {\bf 100} (1976)
310--331.
%%CITATION = APNYA,100,310;%%.

\bibitem{Israel:1979wp}
W.~Israel and J.~M. Stewart, ``{Transient relativistic thermodynamics and
  kinetic theory},'' {\em Ann. Phys.} {\bf 118} (1979)
341--372.
%%CITATION = APNYA,118,341;%%.

\bibitem{Lublinsky:2009kv}
M.~Lublinsky and E.~Shuryak, ``{Improved Hydrodynamics from the AdS/CFT},''
  {\em Phys. Rev.} {\bf D80} (2009) 065026,
\href{http://arXiv.org/abs/0905.4069}{{\tt 0905.4069}}.
%%CITATION = 0905.4069;%%.

\bibitem{Kovtun:2005ev}
P.~K. Kovtun and A.~O. Starinets, ``{Quasinormal modes and holography},'' {\em
  Phys. Rev.} {\bf D72} (2005) 086009,
\href{http://arXiv.org/abs/hep-th/0506184}{{\tt hep-th/0506184}}.
%%CITATION = HEP-TH/0506184;%%.

\bibitem{Berti:2009kk}
E.~Berti, V.~Cardoso, and A.~O. Starinets, ``{Quasinormal modes of black holes
  and black branes},'' {\em Class. Quant. Grav.} {\bf 26} (2009) 163001,
\href{http://arXiv.org/abs/0905.2975}{{\tt 0905.2975}}.
%%CITATION = 0905.2975;%%.

\bibitem{Arfken}
G.~Arfken and H.~Weber, {\em {Mathematical methods for physicists}}.
\newblock Elsevier, 2005.

\bibitem{Danielsson:1998wt}
U.~H. Danielsson, E.~Keski-Vakkuri, and M.~Kruczenski, ``{Vacua, Propagators,
  and Holographic Probes in AdS/CFT},'' {\em JHEP} {\bf 01} (1999) 002,
\href{http://arXiv.org/abs/hep-th/9812007}{{\tt hep-th/9812007}}.
%%CITATION = HEP-TH/9812007;%%.

\bibitem{Witten:1998zw}
E.~Witten, ``{Anti-de Sitter space, thermal phase transition, and confinement
  in gauge theories},'' {\em Adv. Theor. Math. Phys.} {\bf 2} (1998) 505--532,
\href{http://arXiv.org/abs/hep-th/9803131}{{\tt hep-th/9803131}}.
%%CITATION = HEP-TH/9803131;%%.

\bibitem{Brower:2000rp}
R.~C. Brower, S.~D. Mathur, and C.-I. Tan, ``{Glueball Spectrum for QCD from
  AdS Supergravity Duality},'' {\em Nucl. Phys.} {\bf B587} (2000) 249--276,
\href{http://arXiv.org/abs/hep-th/0003115}{{\tt hep-th/0003115}}.
%%CITATION = HEP-TH/0003115;%%.

\bibitem{Bigazzi:2009bk}
F.~Bigazzi {\em et al.}, ``{D3-D7 Quark-Gluon Plasmas},'' {\em JHEP} {\bf 11}
  (2009) 117,
\href{http://arXiv.org/abs/0909.2865}{{\tt 0909.2865}}.
%%CITATION = 0909.2865;%%.

\bibitem{Abelev:2008nda}
{\bf STAR} Collaboration, B.~I. Abelev {\em et al.}, ``{Indications of Conical
  Emission of Charged Hadrons at RHIC},'' {\em Phys. Rev. Lett.} {\bf 102}
  (2009) 052302,
\href{http://arXiv.org/abs/0805.0622}{{\tt 0805.0622}}.
%%CITATION = 0805.0622;%%.

\bibitem{deHaro:2000xn}
S.~de~Haro, S.~N. Solodukhin, and K.~Skenderis, ``{Holographic reconstruction
  of spacetime and renormalization in the AdS/CFT correspondence},'' {\em
  Commun. Math. Phys.} {\bf 217} (2001) 595--622,
\href{http://arXiv.org/abs/hep-th/0002230}{{\tt hep-th/0002230}}.
%%CITATION = HEP-TH/0002230;%%.

\end{thebibliography}\endgroup
\end{document}